\documentclass[12pt]{article}
\usepackage{amssymb}
\renewcommand{\theequation}{\arabic{section}.\arabic{equation}}

\begin{document}



\def\a{\alpha}
\def\b{\beta}
\def\d{\delta}
\def\e{\epsilon}
\def\g{\gamma}
\def\h{\mathfrak{h}}
\def\k{\kappa}
\def\l{\lambda}
\def\o{\omega}
\def\p{\wp}
\def\r{\rho}
\def\t{\theta}
\def\s{\sigma}
\def\z{\zeta}
\def\x{\xi}
 \def\A{{\cal{A}}}
 \def\B{{\cal{B}}}
 \def\C{{\cal{C}}}
 \def\D{{\cal{D}}}
\def\G{\Gamma}
\def\K{{\cal{K}}}
\def\O{\Omega}
\def\L{\Lambda}
\def\f{E_{\tau,\eta}(sl_2)}
\def\E{E_{\tau,\eta}(sl_n)}
\def\T{{\cal{T}}}
\def\Zb{\mathbb{Z}}
\def\Cb{\mathbb{C}}

\def\R{\overline{R}}

\def\beq{\begin{equation}}
\def\eeq{\end{equation}}
\def\bea{\begin{eqnarray}}
\def\eea{\end{eqnarray}}
\def\ba{\begin{array}}
\def\ea{\end{array}}
\def\no{\nonumber}
\def\le{\langle}
\def\re{\rangle}
\def\lt{\left}
\def\rt{\right}

\newtheorem{Theorem}{Theorem}
\newtheorem{Definition}{Definition}
\newtheorem{Proposition}{Proposition}
\newtheorem{Lemma}{Lemma}
\newtheorem{Corollary}{Corollary}
\newcommand{\proof}[1]{{\bf Proof. }
        #1\begin{flushright}$\Box$\end{flushright}}

\baselineskip=20pt

\newfont{\elevenmib}{cmmib10 scaled\magstep1}
\newcommand{\preprint}{
   \begin{flushleft}
     \elevenmib Yukawa\, Institute\, Kyoto\\
   \end{flushleft}\vspace{-1.3cm}
   \begin{flushright}\normalsize  \sf
     YITP-03-56\\
     {\tt hep-th/0308127} \\ August 2003
   \end{flushright}}
\newcommand{\Title}[1]{{\baselineskip=26pt
   \begin{center} \Large \bf #1 \\ \ \\ \end{center}}}
\newcommand{\Author}{\begin{center}
   \large \bf W.\,-L. Yang${}^{a,b}$~ and~R.~Sasaki${}^b$ \end{center}}
\newcommand{\Address}{\begin{center}

     ${}^a$ Institute of Modern Physics, Northwest University\\
     Xian 710069, P.R. China\\
     ~~\\
     $^b$ Yukawa Institute for Theoretical Physics,\\
     Kyoto University, Kyoto 606-8502, Japan
   \end{center}}
\newcommand{\Accepted}[1]{\begin{center}
   {\large \sf #1}\\ \vspace{1mm}{\small \sf Accepted for Publication}
   \end{center}}

\preprint
\thispagestyle{empty}
\bigskip\bigskip\bigskip

\Title{Exact solution of $\Zb_n$ Belavin model with open boundary
condition} \Author

\Address
\vspace{1cm}

\begin{abstract}
$\Zb_n$ Belavin model with open boundary condition is studied. The
{\it double-row transfer matrices\/} of the model are diagonalized
by algebraic Bethe ansatz method in terms of the intertwiner and
the face-vertex correspondence relation. The eigenvalues and the
corresponding Bethe ansatz equations are obtained.

\vspace{1truecm} \noindent {\it PACS:} 03.65.Fd; 75.10.Jm;
05.30.-d

\end{abstract}
\newpage
\section{Introduction}
\label{intro} \setcounter{equation}{0}

Exactly solvable two-dimensional lattice models have been
attracting a great deal of interest from both physicists and
mathematician. Bethe ansatz method has been proved to be the most
powerful and (probably) unified tool to construct the common
eigenvectors of commuting families of operators (so-called {\it
transfer matrices\/}) for various models
\cite{Tak79,Tha82,Bax82,Kor93}. Two-dimensional exactly solvable
lattice models have traditionally been solved by imposing periodic
boundary condition. The Yang-Baxter equation \cite{Yan67,Bax82}
\bea R_{12}(u_1-u_2)R_{13}(u_1-u_3)R_{23}(u_2-u_3)=
R_{23}(u_2-u_3)R_{13}(u_1-u_3)R_{12}(u_1-u_2), \label{YBE-V} \eea
together with such boundary condition then leads to families of
commuting {\it one-row transfer matrices\/} and hence solvability
\cite{Bax82}. The work of Sklyanin \cite{Skl88} shows that, by
using the reflection equation (RE) introduced by Cherednik
\cite{Che84}
 \bea R_{12}(u_1-u_2)K_1(u_1)R_{21}(u_1+u_2)K_2(u_2)=
K_2(u_2)R_{12}(u_1+u_2)K_1(u_1)R_{21}(u_1-u_2),\label{RE-V1}\eea
it is also possible to construct families of commuting {\it
double-row transfer matrices} for the models with open boundary
condition \cite{Skl88,Veg89,Mez91}. Both the {\it one-row\/} (for
periodic boundary condition) and the {\it double-row\/} (for open
boundary condition) {\it transfer matrices} for rational and
trigonometric integrable models can be very successfully
diagonalized by algebraic Bethe ansatz method
\cite{Bab82,Sch83,Veg94}.

In contrast to the rational and trigonometric models, the elliptic
models of vertex type, such as the eight-vertex model (or $\Zb_2$
Belavin model) and $\Zb_n$ Belavin model \cite{Bel81}, had been
problematic due to the fact that  the corresponding pseudo-vacuum
state could not be constructed directly in vertex form
\cite{Bax82,Fan96,Hou89,Hou03}. Or from the representation
theories point of view, the highest weight representations of the
underlying algebras---$\Zb_n$ Skyanin algebras \cite{Skl83,Hou891}
were not properly defined. Therefore, the algebraic Bethe ansatz
method had not been applied to the elliptic integrable models
directly.

It is well-known that in Baxter's original work \cite{Bax71} for
the eight-vertex model with periodic boundary condition, he
elegantly used the intertwiner  to transform the eight-vertex
model (vertex type) to $A^{(1)}_1$ solid-on-solid (SOS) model
(face type) through the face-vertex correspondence relation, then
he succeeded in constructing the corresponding pseudo-vacuum state
and in diagonalizing the transfer matrices by algebraic Bethe
ansatz method. This method was later generalized to diagonalize
$\Zb_n$ Belavin model with periodic boundary condition
\cite{Hou89} by using the generalized intertwiner  and the
face-vertex correspondence relation given by \cite{Jim87}. In
\cite{Fan96}, it was further generalized to diagonalize the
eight-vertex model with open boundary condition.

In this paper, we will extend the above construction of Bethe
ansatz method to the generic $\Zb_n$ Belavin model with open
boundary condition. In section 2, we first review $\Zb_n$ Belavin
model and reflecting open boundary condition; the model under
consideration in this paper is constructed from the corresponding
{\it double-row transfer matrices\/}. In section 3, we introduce
the intertwiner vectors and accompanying face-vertex
correspondence relations which will play key roles in transforming
the model in ``vertex language" to the one in the ``face
language". After finding the pseudo-vacuum state, we use the
algebraic Bethe ansatz method to diagonalize the transfer matrices
of the model in section 4. Section 5 is for conclusions. Some
detailed technical calculations are given in appendix A-C.

\section{$\Zb_n$ Belavin model and integrable boundary condition}
 \label{Vertex} \setcounter{equation}{0}

\subsection{$\Zb_n$ Belavin R-matrix}

Let us fix $\tau$ such that $Im(\tau)>0$ and a generic complex
number $w$. Introduce the following elliptic functions \bea
&&\t\lt[
\begin{array}{c}
a\\b
\end{array}\rt](u,\tau)=\sum_{m=-\infty}^{\infty}
exp\lt\{\sqrt{-1}\pi\lt[(m+a)^2\tau+2(m+a)(u+b)\rt]\rt\},\\
&&\t^{(j)}(u)=\t\lt[\begin{array}{c}\frac{1}{2}-\frac{j}{n}\\
[2pt]\frac{1}{2}
\end{array}\rt](u,n\tau),~~~
\s(u)=\t\lt[\begin{array}{c}\frac{1}{2}\\[2pt]\frac{1}{2}
\end{array}\rt](u,\tau).\label{Function}
\eea Among them the $\s$-function\footnote{Our $\s$-function is
the $\vartheta$-function $\vartheta_1(u)$ \cite{Whi50}. It has the
following relation with the {\it Weierstrassian\/} $\s$-function
if denoted it by $\s_w(u)$: $\s_w(u)\propto e^{\eta_1u^2}\s(u)$,
$\eta_1=\pi^2(\frac{1}{6}-4\sum_{n=1}^{\infty}\frac{nq^{2n}}{1-q^{2n}})
$ and $q=e^{\sqrt{-1}\tau}$.}
 satisfies the following
identity:\bea
&&\s(u+x)\s(u-x)\s(v+y)\s(v-y)-\s(u+y)\s(u-y)\s(v+x)\s(v-x)\no\\
&&~~~~~~=\s(u+v)\s(u-v)\s(x+y)\s(x-y),\label{identity}\eea which
will be useful in deriving equations in the following.

The $\Zb_n$ Belavin R-matrix \cite{Bel81} is given by \cite{Jim87}
\bea R^B(u)=\sum_{i,j,k,l}R^{kl}_{ij}(u)E_{ik}\otimes
E_{lj},\label{Belavin-R}\eea in which $E_{ij}$ is the matrix with
elements $(E_{ij})^l_k=\d_{jk}\d_{il}$. The coefficient functions
are \bea R^{kl}_{ij}(u)=\lt\{
\begin{array}{ll}
\frac{h(u)\s(w)\t^{(i-j)}(u+w)}{\s(u+w)\t^{(i-k)}(w)\t^{(k-j)}(u)}&{\rm
if}~i+j=k+l~mod~n,\\[6pt]
0&{\rm otherwise}.
\end{array}\rt.\label{Belavin-R1}
\eea Here we have set \bea
h(u)=\frac{\prod_{j=0}^{n-1}\t^{(j)}(u)}
{\prod_{j=1}^{n-1}\t^{(j)}(0)}.\label{Belavin-R2}\eea The R-matrix
satisfies the quantum Yang-Baxter equation (\ref{YBE-V})
 and the following
unitarity and crossing-unitarity relations \cite{Ric86} \bea
&&{\rm
Unitarity}:~~{R^B}_{12}(u)R^B_{21}(-u)= id,\label{Unitarity}\\
&&\mbox{
Crossing-unitarity}:~~(R^B)^{t_2}_{21}(-u-nw)(R^B)_{12}^{t_2}(u)
=\frac{e^{\sqrt{-1}nw}\s(u)\s(u+nw)}{\s(u+w)\s(u+nw-w)}~id,\no\\
&&~~\label{crosing-unitarity} \eea where $t_i,~i=1,2$ denotes
transposition in the $i$-th space.

We introduce  ``row-to-row" monodromy matrix $T(u)$, which is an
$n\times n$ matrix with elements being operators acting  on
$(\Cb^n)^{\otimes N}$, from the R-matrix by the standard way
\cite{Kor93}  \bea T_0(u)=R^B_{01}(u+z_1)R^B_{02}(u+z_2)\cdots
R^B_{0N}(u+z_N).\label{T-matrix}\eea Here $\{z_i|i=1,\cdots, N\}$
are arbitrary free complex parameters which are usually called
inhomogeneous parameters. One can show that $T(u)$ satisfies the
so-called ``RLL" relation from the Yang-Baxter equation
(\ref{YBE-V}) \bea
R^B_{12}(u-v)T_1(u)T_2(v)=T_2(v)T_1(u)R^B_{12}(u-v).\label{Relation1}\eea
\subsection{Integrable boundary condition}
We proceed to study $\Zb_n$ Belavin model with open boundary, by
following Sklyanin's work \cite{Skl88}. Let us introduce a pair of
K-matrices $K^-(u)$ and $K^+(u)$. The former $K^-(u)$ satisfies RE
(\ref{RE-V1}), namely,
 \bea &&R^B_{12}(u_1-u_2)K^-_1(u_1)R^B_{21}(u_1+u_2)K^-_2(u_2)\no\\
 &&~~~~~~=
K^-_2(u_2)R^B_{12}(u_1+u_2)K^-_1(u_1)R^B_{21}(u_1-u_2),\label{RE-V}\eea
and the latter  $K^+(u)$ satisfies its dual equation \bea
&&R^B_{12}(u_2-u_1)K^+_1(u_1)R^B_{21}(-u_1-u_2-nw)K^+_2(u_2)\no\\
&&~~~~~~=
K^+_2(u_2)R^B_{12}(-u_1-u_2-nw)K^+_1(u_1)R^B_{21}(u_2-u_1).
\label{DRE-V}\eea Various integrable boundary conditions are
described by different solutions of $K^-(u)$ and $K^+(u)$
\cite{Skl88,Gho94}. In this paper, we shall consider the solution
of RE (\ref{RE-V}) $K^-(u)$ given by \cite{Fan97}\bea
K^-(u)^s_t=\sum_{i=1}^n
\frac{\s(\l^{(-)}_iw+\xi-u)}{\s(\l^{(-)}_iw+\xi+u)}
\phi^{(s)}_{\l^{(-)},\l^{(-)}-\hat{i}}(u)
\bar{\phi}^{(t)}_{\l^{(-)},\l^{(-)}-\hat{i}}(-u),
\label{K-matrix}\eea and the solution of the dual RE (\ref{DRE-V})
given by \cite{Sak03} \bea &&K^+(u)^s_t= \sum_{i=1}^n
\lt\{\prod_{k\neq
i}\frac{\s((\l^{(+)}_i-\l^{(+)}_k)w-w)}{\s((\l^{(+)}_i-\l^{(+)}_k)w)}\rt\}
\frac{\s(\l^{(+)}_iw+\bar{\xi}+u+
\frac{nw}{2})}{\s(\l^{(+)}_iw+\bar{\xi}-u-\frac{nw}{2})}\no\\
&&~~~~~~~~~~~~~~~~~~~\times
\phi^{(s)}_{\l^{(+)},\l^{(+)}-\hat{i}}(-u)
\tilde{\phi}^{(t)}_{\l^{(+)},\l^{(+)}-\hat{i}}(u).\label{K-matrix1}\eea
They depend on  free parameters $\{\l^{(-)}_i|i=1,\cdots,n\}$ and
$\xi$ (resp. $\{\l^{(+)}_i|i=1,\cdots,n\}$ and $\bar{\xi}$) which
specify the left integrable boundary condition (resp. the right
integrable boundary condition). It is very convenient to introduce
two vectors $\l^{(\pm)}=\sum_{i}\l^{(\pm)}_i\e_i$ associated with
the boundary parameters $\{\l^{(\pm)}_i\}$, where
$\{\e_i,~i=1,\cdots,n\}$ is the orthonormal basis of the vector
space $\Cb^n$ such that $\langle \e_i,\e_j\rangle=\d_{ij}$. In
equations (\ref{K-matrix}), (\ref{K-matrix1})
$\phi,\,\bar{\phi},\,\tilde{\phi}$ are intertwiners which will be
specified later in section 3. We consider the  generic
$\{\l^{(\pm)}_i\}$ such that $\l^{(\pm)}_iw\neq \l^{(\pm)}_jw\,\,
(modulo ~\Zb+\tau\Zb)$ for $i\neq j$. This condition is necessary
for the non-singularity of $K^+(u)$. Let us remark that a further
restriction  \bea \l^{(+)}=\l^{(-)}=\l_0,~~{\rm a~generic~vector}~
\l_0\in \Cb^n,\label{boundary-res}\eea is necessary for the
application of the algebraic Bethe ansatz method in section 4.
Hereafter we will consider only the above case.

For open boundary condition case, instead of the standard
``row-to-row" monodromy matrix $T(u)$ (\ref{T-matrix}), one needs
to introduce a ``double-row" monodromy matrix $\mathbb{T}(u)$ \bea
\mathbb{T}(u)=T(u)K^-(u)T^{-1}(-u).\label{Mon-V}\eea Using the
``RLL" relation (\ref{Relation1}) and RE (\ref{RE-V}), one can
prove that $\mathbb{T}(u)$ satisfies the reflection equation
 \bea R^B_{12}(u_1-u_2)\mathbb{T}_1(u_1)R^B_{21}(u_1+u_2)
 \mathbb{T}_2(u_2)=
\mathbb{T}_2(u_2)R^B_{12}(u_1+u_2)\mathbb{T}_1(u_1)R^B_{21}(u_1-u_2).
\label{Relation-Re}\eea The {\it double-row transfer matrices\/}
of  $\Zb_n$ Belavin model with open boundary are given by: \bea
t(u;\xi)=tr(K^+(u)\mathbb{T}(u)).\label{trans}\eea Here we have
emphasized the dependence on the boundary parameter $\xi$ of the
transfer matrices through the boundary K-matrix $K^-(u)$
(\ref{K-matrix})\footnote{It will be  shown in section 4 that the
spectral parameter $u$ and the boundary parameter $\xi$ will be
shifted for the reduced transfer matrices in each step of the
nested Bethe anstz procedure. Therefore, it is convenient to
specify the dependence on the boundary parameter $\xi$ in addition
to the spectral parameter $u$. }. With the help of the unitarity
(\ref{Unitarity}) and crossing-unitarity relations
(\ref{crosing-unitarity}) of R-matrix, Yang-Baxter relation
(\ref{YBE-V}), the RE (\ref{RE-V}) and its dual (\ref{DRE-V}), we
can prove that the transfer matrices with different spectral
parameters  commute with each other \cite{Skl88}:
$[t(u;\xi),t(v;\xi)]=0$. This ensures the integrability of the
system. The aim of this paper is to find the common eigenvalues
and eigenvectors of the transfer matrices (\ref{trans}) with the
special K-matrices $K^{\pm}(u)$ given by (\ref{K-matrix}),
(\ref{K-matrix1}) and (\ref{boundary-res}).


\section{$A^{(1)}_{n-1}$ SOS R-matrix and face-vertex correspondence}
 \label{Face} \setcounter{equation}{0}

The $A_{n-1}$ simple roots are
$\lt\{\a_{i}=\e_i-\e_{i+1}~|~i=1,\cdots,n-1\rt\}$ and the
fundamental weights $\lt\{\L_i~|~i=1,\cdots,n-1\rt\}$ satisfying
$\langle\L_i,~\a_j\rangle=\d_{ij}$ are given by \bea
\L_i=\sum_{k=1}^{i}\e_k-\frac{i}{n}\sum_{k=1}^{n}\e_k. \no\eea Set
\bea \hat{i}=\e_i-\overline{\e},~~\overline{\e}=
\frac{1}{n}\sum_{k=1}^{n}\e_k,~~i=1,\cdots,n,~~{\rm
then}~~\sum_{i=1}^n\hat{i}=0. \label{Vectors} \eea For each
dominant weight $\L=\sum_{i=1}^{n-1}a_i\L_{i}~,~~a_{i}\in
\Zb^+$(the set of non-negative integers), there exists an
irreducible highest weight finite-dimensional representation
$V_{\L}$ of $A_{n-1}$ with the highest vector $ |\L\rangle$. For
example the fundamental vector representation is $V_{\L_1}$.

Let $\h$ be the Cartan subalgebra of $A_{n-1}$ and $\h^{*}$ be its
dual. A finite dimensional diagonalisable  $\h$-module is a
complex finite dimensional vector space $W$ with a weight
decomposition $W=\oplus_{\mu\in \h^*}W[\mu]$, so that $\h$ acts on
$W[\mu]$ by $x~v=\mu(x)~v$, $(x\in \h,~~v\in~W[\mu])$. For
example, the fundamental vector representation $V_{\L_1}=\Cb^n$,
the non-zero weight spaces $W[\hat{i}]=\Cb \e_i,~i=1,\cdots,n$.

For a generic $\l\in \Cb^n$, define \bea
\l_i=\langle\l,\e_i\rangle,
~~\l_{ij}=\l_i-\l_j=\langle\l,\e_i-\e_j\rangle,~~i,j=1,\cdots,n.
\label{Def1}\eea Let $R(u,\l)\in End(\Cb^n\otimes\Cb^n)$ be the
R-matrix of the $A^{(1)}_{n-1}$ SOS model \cite{Jim87} given by
\bea &&R(u,\l)=\sum_{i=1}^{n}R^{ii}_{ii}(u,\l)E_{ii}\otimes E_{ii}
+\sum_{i\ne j}\lt\{R^{ij}_{ij}(u,\l)E_{ii}\otimes E_{jj}+
R^{ji}_{ij}(u,\l)E_{ji}\otimes E_{ij}\rt\}.\no\\
\label{R-matrix} \eea The coefficient functions are \bea
&&R^{ii}_{ii}(u,\l)=1,~~
R^{ij}_{ij}(u,\l)=\frac{\s(u)\s(\l_{ij}w-w)}
{\s(u+w)\s(\l_{ij}w)},\label{Elements1}\\
&& R^{ji}_{ij}(u,\l)=\frac{\s(w)\s(u+\l_{ij}w)}
{\s(u+w)\s(\l_{ij}w)},\label{Elements2}\eea  and  $\l_{ij}$ is
defined in (\ref{Def1}). The R-matrix satisfies the dynamical
(modified) quantum Yang-Baxter equation \bea
&&R_{12}(u_1-u_2,\l-h^{(3)})R_{13}(u_1-u_3,\l)
R_{23}(u_2-u_3,\l-h^{(1)})\no\\
&&~~~~=R_{23}(u_2-u_3,\l)R_{13}(u_1-u_3,\l-h^{(2)})R_{12}(u_1-u_2,\l),
\label{MYBE}\eea with the initial condition \bea
R^{kl}_{ij}(0,\l)=\d^l_i~\d^k_j.\label{Initial}\eea We adopt the
notation: $R_{12}(u,\l-h^{(3)})$ acts on a tensor $v_1\otimes v_2
\otimes v_3$ as $R(u,\l-\mu)\otimes id$ if $v_3\in W[\mu]$.
Moreover, the R-matrix satisfies unitarity and a modified
crossing-unitarity relation \cite{Jim87,Fan97}.

Let us introduce an intertwiner---an $n$-component  column vector
$\phi_{\l,\l-\hat{j}}(u)$ whose  $k$-th element is \bea
\phi^{(k)}_{\l,\l-\hat{j}}(u)=\t^{(k)}(u+nw\l_j).\label{Intvect}\eea
Using the intertwiner, the face-vertex correspondence can be
written as \cite{Jim87}\bea R^B_{12}(u_1-u_2)
\phi_{\l,\l-\hat{i}}(u_1)\otimes
\phi_{\l-\hat{i},\l-\hat{i}-\hat{j}}(u_2)=
\sum_{k,l}R(u_1-u_2,\l)^{kl}_{ij}
\phi_{\l-\hat{l},\l-\hat{l}-\hat{k}}(u_1)\otimes
\phi_{\l,\l-\hat{l}}(u_2).\no\\
\label{Face-vertex}\eea Then the Yang-Baxter equation of $\Zb_n$
Belavin's R-matrix $R^B(u)$ (\ref{YBE-V}) is equivalent to the
dynamical Yang-Baxter equation of $A^{(1)}_{n-1}$ SOS R-matrix
$R(u,\l)$ (\ref{MYBE}). For a generic $\l$, we can introduce other
types of intertwiners $\bar{\phi},~\tilde{\phi}$  satisfying the
following conditions \bea
&&\sum_{k=1}^n\bar{\phi}^{(k)}_{\l,\l-\hat{\mu}}(u)
~\phi^{(k)}_{\l,\l-\hat{\nu}}(u)=\d_{\mu\nu},\label{Int1}\\
&&\sum_{k=1}^n\tilde{\phi}^{(k)}_{\l+\hat{\mu},\l}(u)
~\phi^{(k)}_{\l+\hat{\nu},\l}(u)=\d_{\mu\nu}.\label{Int2}\eea One
can derive the following relations from the above conditions \bea
&&\sum_{\mu=1}^n\bar{\phi}^{(i)}_{\l,\l-\hat{\mu}}(u)
~\phi^{(j)}_{\l,\l-\hat{\mu}}(u)=\d_{ij},\label{Int3}\\
&&\sum_{\mu=1}^n\tilde{\phi}^{(i)}_{\l+\hat{\mu},\l}(u)
~\phi^{(j)}_{\l+\hat{\mu},\l}(u)=\d_{ij}.\label{Int4}\eea With the
help of the properties of $\phi,\,\bar{\phi},\,\tilde{\phi}$
(\ref{Int1})-(\ref{Int4}), we can derive the following relations
from the face-vertex correspondence relation (\ref{Face-vertex})
\bea &&(\tilde{\phi}_{\l+\hat{k},\l}(u_1)\otimes
1)R^B_{12}(u_1-u_2) (1\otimes\phi_{\l+\hat{j},\l}(u_2))\no\\
&&~~~~~~~~= \sum_{i,l}R(u_1-u_2,\l)^{kl}_{ij}
\tilde{\phi}_{\l+\hat{i}+\hat{j},\l+\hat{j}}(u_1)\otimes
\phi_{\l+\hat{k}+\hat{l},\l+\hat{k}}(u_2),\label{Face-vertex1}\\
&&(\tilde{\phi}_{\l+\hat{k},\l}(u_1)\otimes
\tilde{\phi}_{\l+\hat{k}+\hat{l},\l+\hat{k}}(u_2))R^B_{12}(u_1-u_2)\no\\
&&~~~~~~~~= \sum_{i,j}R(u_1-u_2,\l)^{kl}_{ij}
\tilde{\phi}_{\l+\hat{i}+\hat{j},\l+\hat{j}}(u_1)\otimes
\tilde{\phi}_{\l+\hat{j},\l}(u_2),\label{Face-vertex2}\\
&&(1\otimes \bar{\phi}_{\l,\l-\hat{l}}(u_2))R^B_{12}(u_1-u_2)
(\phi_{\l,\l-\hat{i}}(u_1)\otimes 1)\no\\
&&~~~~~~~~= \sum_{k,j}R(u_1-u_2,\l)^{kl}_{ij}
\phi_{\l-\hat{l},\l-\hat{k}-\hat{l}}(u_1)\otimes
\bar{\phi}_{\l-\hat{i},\l-\hat{i}-\hat{j}}(u_2),\label{Face-vertex3}\\
&&(\bar{\phi}_{\l-\hat{l},\l-\hat{k}-\hat{l}}(u_1)\otimes
\bar{\phi}_{\l,\l-\hat{l}}(u_2))R^B_{12}(u_1-u_2)\no\\
&&~~~~~~~~= \sum_{i,j}R(u_1-u_2,\l)^{kl}_{ij}
\bar{\phi}_{\l,\l-\hat{i}}(u_1)\otimes
\bar{\phi}_{\l-\hat{i},\l-\hat{i}-\hat{j}}(u_2).\label{Face-vertex4}
\eea

The face-vertex correspondence relations (\ref{Face-vertex}) and
(\ref{Face-vertex1})-(\ref{Face-vertex4}) will play an important
role to {\it translate\/} all formulas in ``vertex language" into
their ``face language" form so that the algebraic Bethe ansatz
method can be applied to diagonize the transfer matrices.

\section{Algebraic Bethe ansatz for $\Zb_n$ Belavin model
with open boundary condition}
\label{BAE}
\setcounter{equation}{0}

As mentioned in Introduction, the intertwiners and the face-vertex
correspondence relations (\ref{Intvect})-(\ref{Face-vertex4}) will
play a fundamental role in the construction of the eigenstates of
$\Zb_n$ Belavin model with open boundary condition specified by
the K-matrices $K^{\pm}(u)$ given in (\ref{K-matrix}),
(\ref{K-matrix1}) and (\ref{boundary-res}). In order to apply the
algebraic Bethe ansatz method, we need to transform the
fundamental exchange relation (\ref{Relation-Re}) of vertex type
into its face type so that we can construct the corresponding
pseudo-vacuum and the creation operators to construct associated
Bethe ansatz states.

\subsection{Exchange relations of double-row monodromy
matrix of face type}

The transfer matrices of  $\Zb_n$ Belavin model with open boundary
condition (\ref{trans}) can be rewritten in the face type form by
using (\ref{Int3}) and (\ref{Int4}) \bea
&&t(u;\xi)=tr(K^+(u)\mathbb{T}(u))\no\\
&&~~=\sum_{\mu,\nu}tr\lt(K^+(u) \phi_{\l-\hat{\mu}+\hat{\nu},
\l-\hat{\mu}}(u)\tilde{\phi}_{\l-\hat{\mu}+\hat{\nu},
\l-\hat{\mu}}(u)~\mathbb{T}(u) \phi_{\l, \l-\hat{\mu}}(-u)
\bar{\phi}_{\l, \l-\hat{\mu}}(-u)\rt)\no\\
&&~~=\sum_{\mu,\nu}\bar{\phi}_{\l, \l-\hat{\mu}}(-u)K^+(u)
\phi_{\l-\hat{\mu}+\hat{\nu},\l-\hat{\mu}}(u)~
\tilde{\phi}_{\l-\hat{\mu}+\hat{\nu},
\l-\hat{\mu}}(u)~\mathbb{T}(u)\phi_{\l, \l-\hat{\mu}}(-u)\no\\
&&~~=\sum_{\mu,\nu}\tilde{\K}(\l|u)_{\nu}^{\mu}\T(\l|u;\xi)^{\nu}_{\mu}.
\label{De1}\eea We have introduced the dual K-matrix
$\tilde{\K}(\l|u)$ of face type \cite{Sak03} and the double-row
monodromy matrix $\T(\l|u;\xi)$ of face type as follows \bea
&&\tilde{\K}(\l|u)_{\nu}^{\mu}=\bar{\phi}_{\l,
\l-\hat{\mu}}(-u)K^+(u)
\phi_{\l-\hat{\mu}+\hat{\nu},\l-\hat{\mu}}(u)
\equiv\sum_{i,j}\bar{\phi}^{(j)}_{\l, \l-\hat{\mu}}(-u)K^+(u)^j_i
\phi^{(i)}_{\l-\hat{\mu}+\hat{\nu},\l-\hat{\mu}}(u),\no\\
\label{F-V1}\\
&&\T(\l|u;\xi)^{\nu}_{\mu}=\tilde{\phi}_{\l-\hat{\mu}+\hat{\nu},
\l-\hat{\mu}}(u)~\mathbb{T}(u)\phi_{\l, \l-\hat{\mu}}(-u)\equiv
\sum_{i,j}\tilde{\phi}^{(j)}_{\l-\hat{\mu}+\hat{\nu},
\l-\hat{\mu}}(u)~\mathbb{T}(u)^j_i\phi^{(i)}_{\l,
\l-\hat{\mu}}(-u).\no\\\label{Mon-F} \eea The dependence on the
boundary parameter $\xi$ of $\T(\l|u;\xi)$ is through the
double-row monodromy matrix of vertex type $\mathbb{T}(u)$ in
(\ref{Mon-F}) by the definitions (\ref{Mon-V}) and
(\ref{K-matrix}). One can derive the following exchange relations
among $\T(\l|u;\xi)^{\nu}_{\mu}$ from the exchange relation
(\ref{Relation-Re}), the face-vertex correspondence relation
(\ref{Face-vertex}) and the relation (\ref{Int4}) (for details,
see Appendix A) \bea &&\sum_{i_1,i_2}\sum_{j_1,j_2}~
R(u_1-u_2,\l)^{i_0,j_0}_{i_1,j_1}\T(\l+\hat{j_1}+\hat{i_2}|u_1;\xi)
^{i_1}_{i_2}R(u_1+u_2,\l)^{j_1,i_2}_{j_2,i_3}
\T(\l+\hat{j_3}+\hat{i_3}|u_2;\xi)^{j_2}_{j_3}\no\\
&&~~=\sum_{i_1,i_2}\sum_{j_1,j_2}~
\T(\l+\hat{j_1}+\hat{i_0}|u_2;\xi)
^{j_0}_{j_1}R(u_1+u_2,\l)^{i_0,j_1}_{i_1,j_2}
\T(\l+\hat{j_2}+\hat{i_2}|u_1;\xi)^{i_1}_{i_2}
R(u_1-u_2,\l)^{j_2,i_2}_{j_3,i_3}.\no\\
&&~~\label{RE-F} \eea

Next, let us introduce a set of standard notions for convenience:
\footnote{The scalar factors in the definitions of the operators
$\B(\l|u)$ and $\D(\l|u)$ are to make the relevant commutation
relations as concise  as (\ref{Rel-1})-(\ref{Rel-3}). } \bea
&&\A(\l|u)=\T(\l|u)^1_1,~~\B_i(\l|u)=
\frac{\s(w)}{\s(\l_{i1}w)}\T(\l|u)^1_i,~i=2,\cdots,n,\label{Def-AB}\\
&&\D^j_i(\l|u)=\frac{\s(\l_{j1}w-\d_{ij}w)}{\s(\l_{i1}w)}
\{\T(\l|u)^j_i-\d^j_iR(2u,\l+\hat{1})^{j\,1}_{1\,j}\A(\l|u)\}
,\no\\
&&~~~~~~i,j=2,\cdots,n. \label{Def-D}\eea After some tedious
calculation, we have found the commutation relations among
$\A(\l|u)$, $\D(\l|u)$ and $\B(\l|u)$ (for details, see Appendix
B). The relevant commutation relations are \bea
&&\A(\l|u)\B_i(\l-\hat{1}+\hat{i}|v)=
\frac{\s(u+v)\s(u-v-w)}{\s(u+v+w)\s(u-v)}
\B_i(\l-\hat{1}+\hat{i}|v)\A(\l-\hat{1}+\hat{i}|u)\no\\
&&~~~~~~~~-\frac{\s(w)\s(2v)}{\s(u-v)\s(2v+w)}
\frac{\s(u-v-\l_{1i}w+w)} {\s(\l_{1i}w-w)}
\B_i(\l-\hat{1}+\hat{i}|u)\A(\l-\hat{1}+\hat{i}|v)\no\\
&&~~~~~~~~-\frac{\s(w)}{\s(u+v+w)}\sum_{\a=2}^{n}
\frac{\s(u+v+\l_{\a1}w+2w)}{\s(\l_{\a1}w+w)}
\B_{\a}(\l-\hat{1}+\hat{\a}|u)
\D^{\a}_i(\l-\hat{1}+\hat{i}|v),\label{Rel-1}\\
&&\D^k_i(\l|u)\B_j(\l+\hat{j}-\hat{1}|v)=
\frac{\s(u-v+w)\s(u+v+2w)}{\s(u-v)\s(u+v+w)}\lt\{
\sum_{\a_1,\a_2,\b_1,\b_2=2}^n
R(u+v+w,\l-\hat{i})^{k\,\,\,\,\b_2}_{\a_2\,\b_1}\rt.
\no\\
&&~~~~~~~~~~~~~~~~~~~~~~~~~~~~~~~~~\times\lt.
R(u-v,\l+\hat{j})^{\b_1\,\a_1}_{j\,\,\,\,i}
\B_{\b_2}(\l-\hat{i}+\hat{k}+\hat{\b}_2-\hat{1}|v)
\D^{\a_2}_{\a_1}(\l-\hat{1}+\hat{j}|u)\rt\}\no\\
&&~~~~~~~~-\frac{\s(w)\s(2u+2w)}{\s(u-v)\s(2u+w)}\lt\{
\sum_{\a,\b=2}^n\frac{\s(u-v+\l_{1\a}w-w)}{\s(\l_{1\a}w-w)}
R(2u+w,\l-\hat{i})^{k\,\b}_{\a\,i}\rt.
\no\\
&&~~~~~~~~~~~~~~~~~~~~~~~~~~~~~~~~~\times\lt.
\B_{\b}(\l-\hat{i}+\hat{k}+\hat{\b}-\hat{1}|u)
\D^{\a}_{j}(\l-\hat{1}+\hat{j}|v)\rt\}\no\\
&&~~~~~~~~+\frac{\s(w)\s(2v)\s(2u+2w)}{\s(u+v+w)\s(2v+w)\s(2u+w)}\lt\{
\sum_{\a=2}^n\frac{\s(u+v+\l_{1j}w)}{\s(\l_{1j}w-w)}
R(2u+w,\l-\hat{i})^{k\,\a}_{j\,i}\rt.
\no\\
&&~~~~~~~~~~~~~~~~~~~~~~~~~~~~~~~~~\times\lt.
\B_{\a}(\l-\hat{i}+\hat{k}+\hat{\a}-\hat{1}|u)
\A(\l-\hat{1}+\hat{j}|v)\rt\},\label{Rel-2}\\
\no\\
&&\B_i(\l+\hat{i}-\hat{1}|u)\B_j(\l+\hat{i}+\hat{j}-2\hat{1}|v)
=\sum_{\a,\b=2}^nR(u-v,\l-2\hat{1})^{\b\,\a}_{j\,i}\no\\
&&~~~~~~~~~~~~~~~~~~~~~~~~~~~~~~~~~\times\B_{\b}(\l+\hat{\b}-\hat{1}|v)
\B_{\a}(\l+\hat{\a}+\hat{\b}-2\hat{1}|u).\label{Rel-3}\eea

\subsection{Pseudo-Vacuum state}
The algebraic Bethe ansatz, in addition to the relevant
commutation relations (\ref{Rel-1})-(\ref{Rel-3}), requires a
pseudo-vacuum state (also called reference state) which is the
common eigenstate of the operators $\A$, $\D^i_i$ and annihilated
by the operators $\C_i$. In contrast to the rational and
trigonometric models, for elliptic models of vertex type such as
the eight-vertex model (or $\Zb_2$ Belavin model) and $\Zb_n$
Belavin model, the corresponding pseudo-vacuum state cannot be
constructed directly in ``vertex language"
\cite{Bax82,Fan96,Hou89,Hou03}. However, the state can be
successfully constructed when one {\it translates\/} it into
``face language" (or equivalently after some local gauge
transformation \cite{Tak79}). By the same way, we will construct
the corresponding pseudo-vacuum state for $\Zb_n$ Belavin model
with open boundary condition specified by the K-matrices
$K^{\pm}(u)$ given in (\ref{K-matrix}), (\ref{K-matrix1}) and
(\ref{boundary-res}).

Before introducing the pseudo-vacuum state, let us introduce  a
generic state in the quantum space by the intertwiner vector
(\ref{Intvect}) \bea |i_1,\cdots,i_N\rangle
_{m_0}^{m}=\phi_{m_0,m_0-\hat{i}_N}^N(-z_N)
\phi_{m_0-\hat{i}_N,m_0-\hat{i}_N-\hat{i}_{N-1}}^{N-1}(-z_{N-1})\cdots
\phi_{m_0-\sum_{k=2}^N\hat{i}_k,m_0-\sum_{k=1}^N\hat{i}_k}^1(-z_1),\no\\
\label{gstate} \eea where the vectors $m_0,m\in\Cb^n$ and
$m=m_0-\sum_{k=1}^N\hat{i}_k$,  the vector $\phi^k=id\otimes
id\cdots\otimes \stackrel{k-th}{\phi}\otimes id\cdots$.

Now let us evaluate the action of the monodromy  matrix $\T$
(\ref{Mon-F}) on the state (\ref{gstate}). Using the definition of
the double-row monodromy matrix $\mathbb{T}$ (\ref{Mon-V}) and
relations (\ref{Int3})-(\ref{Int4}), we can further write $\T$ in
the following form \bea
&&\T(m|u;\xi)^j_i=\tilde{\phi}_{m-\hat{i}+\hat{j},m-\hat{i}}(u)T(u)K^-(u)
T^{-1}(-u)\phi_{m,m-\hat{i}}(-u)\no\\
&&~~~~=\sum_{\mu,\nu}\tilde{\phi}_{m-\hat{i}+\hat{j},m-\hat{i}}(u)T(u)
\phi_{m_0-\hat{\nu}+\hat{\mu},m_0-\hat{\nu}}(u)
\tilde{\phi}_{m_0-\hat{\nu}+\hat{\mu},m_0-\hat{\nu}}(u)K^-(u)\no\\
&&~~~~~~~~~~~~\times \phi_{m_0,m_0-\hat{\nu}}(-u)
\bar{\phi}_{m_0,m_0-\hat{\nu}}(-u)T^{-1}(-u)\phi_{m,m-\hat{i}}(-u)\no\\
&&~~~~=\sum_{\mu,\nu}T(m-\hat{i},m_0-\hat{\nu}|u)^j_{\mu}
\K(m_0|u;\xi)^{\mu}_{\nu}S(m,m_0|u)^{\nu}_i. \label{Decomp} \eea
Here we have introduced  \bea
&&T(m,m_0|u)^j_{\mu}=\tilde{\phi}_{m+\hat{j},m}(u)
T(u)\phi_{m_0+\hat{\mu},m_0}(u),\label{Def-T}\\
&&S(m,m_0|u)^{\mu}_i=\bar{\phi}_{m_0,m_0-\hat{\mu}}(-u)
T^{-1}(-u)\phi_{m,m-\hat{i}}(-u),\label{Def-S}\\
&&\K(m_0|u;\xi)^j_{i}=\tilde{\phi}_{m_0-\hat{i}+\hat{j},m_0-\hat{i}}(u)
K^-(u)\phi_{m_0,m_0-\hat{i}}(-u). \label{F-V2}\eea We  can
evaluate the action of the operator $T(m,m_0|u)$ on the state
$|i_1,\cdots,i_N\rangle^m_{m_0}$ from the definition
(\ref{T-matrix}) and the  face-vertex correspondence relation
(\ref{Face-vertex}) \bea
&&T(m,m_0|u)^j_{\mu}|i_1,\cdots,i_N\rangle_{m_0}^m
=\tilde{\phi}^0_{m+\hat{j},m}(u)R^B_{01}(u+z_1)
\phi^1_{m_0-\sum_{k=2}^N\hat{i}_k,m_0-\sum_{k=1}^N\hat{i}_k}(-z_1)\no\\
&&~~~~~~~~~~\times \cdots
R^B_{0N}(u+z_N)\phi^N_{m_0,m_0-\hat{i}_N}(-z_N)\phi^0_{m_0+\hat{\mu},m_0}(u)\no\\
&&~~~~~~
=\sum_{\b_1,i'_N}\tilde{\phi}^0_{m+\hat{j},m}(u)R^B_{01}(u+z_1)
\phi^1_{m_0-\sum_{k=2}^N\hat{i}_k,m_0-\sum_{k=1}^N\hat{i}_k}(-z_1)\no\\
&&~~~~~~~~~~\times \cdots
R^B_{0\,N-1}(u+z_{N-1})\phi^0_{m_0+\hat{\mu}-\hat{i}'_N,m_0-\hat{i}_N}(u)
\phi^{N-1}_{m_0-\hat{i}_N,m_0-\hat{i}_{N-1}-\hat{i}_N}(-z_{N-1})\no\\
&&~~~~~~~~~~\times
R(u+z_N,m_0-\hat{i}_N)^{\b_1\,i_N'}_{\mu\,\,i_N}
\phi^N_{m_0+\hat{\mu},m_0+\hat{\mu}-\hat{i}'_N}(-z_N)\no\\
&&~~~~~~\vdots\no\\
&&~~~~~~=R(u+z_1,m)^{j\,\,i'_1}_{\b_{N-1}\,i_1}
R(u+z_2,m+\hat{i}_1)^{\b_{N-1}\,i'_2}_{\b_{N-2}\,i_2}\cdots\no\\
&&~~~~~~~~~~\times
R(u+z_N,m_0-\hat{i}_N)^{\b_1\,\,i'_N}_{\mu\,\,\,i_N}
|i'_1,\cdots,i'_N\rangle_{m_0+\hat{\mu}}^{m+\hat{j}}.
\label{T-element}\eea We adopt the convention that the repeated
indices imply summation over $1,2,\cdots n$ in the last equation.
The property of the R-matrix \bea
R(u,\l)^{kl}_{ij}=R(u,\l\pm(\hat{i}+\hat{j}))^{kl}_{ij}=
R(u,\l\pm(\hat{k}+\hat{l}))^{kl}_{ij}, \eea is helpful to derive
the above equation. Noting the unitarity of $R^B$
(\ref{Unitarity}), $T^{-1}(-u)$ can be written \bea
T_0^{-1}(-u)=R^B_{N\,0}(u-z_N)\cdots R^{B}_{1\,0}(u-z_1).\eea Then
we can evaluate the action of the operator $S(m,m_0|u)$ on the
state $|i_1,\cdots,i_N\rangle^m_{m_0}$ from the face-vertex
correspondence relation (\ref{Face-vertex}) as we have done for
$T(m,m_0|u)$:  \bea
&&S(m,m_0|u)^{\mu}_i|i_1,\cdots,i_N\rangle_{m_0}^{m}=
R(u-z_N,m_0)^{i'_N\,\mu}_{i_N\,\a_{N-1}}
R(u-z_{N-1},m_0-\hat{i}_N)^{i'_{N-1}\,\a_{N-1}}_{i_{N-1}\,\a_{N-2}}\cdots
\no\\
&&~~~~~~~~~~~~~~~~~~~~~~~~\times
R(u-z_1,m_0-\sum_{k=2}^N\hat{i}_k)^{i'_1\,\a_1}_{i_1\,\,i}
|i'_1,\cdots,i'_N\rangle_{m_0-\hat{\mu}}^{m-\hat{i}}.
\label{S-element}\eea We also adopt the convention that the
repeated indices imply summation over $1,2,\cdots n$ in the above
equation. Similarly, we  obtain the action of $\T$ on the state
$|i_1,\cdots,i_N\rangle_{m_0}^{m}$ from  the decomposition
relation (\ref{Decomp}) and equations (\ref{T-element}),
(\ref{S-element}) \bea
&&\T(m|u;\xi)^j_i|i_1,\cdots,i_N\rangle_{m_0}^{m}
\equiv\T(m,m_0|u;\xi)^j_i|i_1,\cdots,i_N\rangle_{m_0}^{m}\no\\
&&~~~~~~=T(m-\hat{i},m_0-\hat{\nu}|u)^j_{\mu}
\K(m_0|u;\xi)^{\mu}_{\nu}S(m,m_0|u)^{\nu}_i
|i_1,\cdots,i_N\rangle_{m_0}^{m}\no\\
&&~~~~~~=R(u+z_1,m-\hat{i})^{j\,\,i''_1}_{\b_{N-1}\,i'_1}
R(u+z_2,m-\hat{i}+\hat{i'}_1)^{\b_{N-1}\,i''_2}_{\b_{N-2}\,i'_2}
\cdots\no\\
&&~~~~~~~~~~\times R(u+z_N,m-\hat{i}+\sum_{k=1}^{N-1}
\hat{i'}_k)^{\b_1\,i''_N}_{\mu\,\,i'_N} \K(m_0|u;\xi)^{\mu}_{\nu}
R(u-z_N,m+\sum_{k=1}^{N} \hat{i}_k)^{i'_N\,\nu}_{i_N\,\a_{N-1}}
\no\\
&&~~~~~~~~~~\times R(u-z_{N-1},m+\sum_{k=1}^{N-1}
\hat{i}_k)^{i'_{N-1}\,\a_{N-1}}_{i_{N-1}\,\a_{N-2}} \cdots
R(u-z_1,m+\hat{i}_1)^{i'_1\,\a_1}_{i_1\,i}
|i''_1,\cdots,i''_N\rangle
_{m_0-\hat{\nu}+\hat{\mu}}^{m-\hat{i}+\hat{j}}.\no\\
\label{Element-F}\eea Here we also adopt the convention that the
repeated indices imply summation over $1,2,\cdots n$ in the above
equation.

If one chooses $\l=\l_0$ in equation (\ref{F-V1}) and $m_0=\l_0$
in  equation (\ref{F-V2}), where $\l_0$ is related to the boundary
parameters by (\ref{boundary-res}), the corresponding face type
boundary K-matrices $\K(\l_0|u;\xi)$ and $\tilde{\K}(\l_0|u)$
simultaneously become diagonal ones from equations
(\ref{K-matrix}), (\ref{K-matrix1}) and the restriction
(\ref{boundary-res}) \bea
\K(\l_0|u;\xi)^j_i=\d_i^jk(\l_0|u;\xi)_i,~~
\tilde{\K}(\l_0|u)^j_i=\d_i^j\tilde{k}(\l_0|u)_i.\label{Diag-F}\eea
The functions $k(\l_0|u;\xi)_i,\,\tilde{k}(\l_0|u)_i $ are given
by \bea &&k(\l_0|u;\xi)_i
=\frac{\s((\l_0)_iw+\xi-u)}{\s((\l_0)_iw+\xi+u)},\label{k-def}\\
&&\tilde{k}(\l_0|u)_i=\lt\{\prod_{k\neq
i}\frac{\s((\l_0)_{ik}w-w)}{\s((\l_0)_{ik}w)}\rt\}
\frac{\s((\l_0)_iw+\bar{\xi}+u+
\frac{nw}{2})}{\s((\l_0)_iw+\bar{\xi}-u-\frac{nw}{2})}.
\label{k-def1}\eea The fact that both $\K(\l_0|u;\xi)$ and
$\tilde{\K}(\l_0|u)$ have diagonal form will enable us to
construct the pseudo-vacuum state of the model and apply the
algebraic Bethe ansatz method to diagonalize the {\it double-row
transfer matrices} (\ref{trans}) later.

Now, let us construct the pseudo-vacuum state $|\O\rangle$: \bea
|\O\rangle\equiv|vac\rangle^{\l_0-N\hat{1}}_{\l_0}=|1,\cdots,1\rangle
^{\l_0-N\hat{1}}_{\l_0},\label{Vac}\eea where $\l_0$ is related to
the boundary parameters of the boundary K-matrices $\K^{\pm}$ in
(\ref{boundary-res}). Then we find that the actions of the
operators $T(\l_0-N\hat{1},\l_0|u)$ given by (\ref{Def-T}) and
$S(\l_0-N\hat{1},\l_0|u)$ given by (\ref{Def-S}) on the
pseudo-vacuum state (\ref{Vac}) become \bea
&&T(\l_0-N\hat{1},\l_0|u)^1_1|vac\rangle^{\l_0-N\hat{1}}_{\l_0}
=|vac\rangle^{\l_0-N\hat{1}+\hat{1}}_{\l_0+\hat{1}},\label{Action-1}\\
&&T(\l_0-N\hat{1},\l_0|u)^i_1|vac\rangle^{\l_0-N\hat{1}}_{\l_0}=0,
~~i=2,\cdots,n,\\
&&T(\l_0-N\hat{1},\l_0|u)^i_j|vac\rangle^{\l_0-N\hat{1}}_{\l_0}
=\d^i_j\prod_{k=1}^NR(u+z_k,\l_0+\hat{j}-(N-k)\hat{1})^{j1}_{j1}
|vac\rangle^{\l_0-N\hat{1}+\hat{j}}_{\l_0+\hat{j}},\no\\
&&~~~~~~~~~~~~~~~~i,j=2,\cdots,n,\\
&&S(\l_0-N\hat{1},\l_0|u)^1_1|vac\rangle^{\l_0-N\hat{1}}_{\l_0}
=|vac\rangle^{\l_0-N\hat{1}-\hat{1}}_{\l_0-\hat{1}},\\
&&S(\l_0-N\hat{1},\l_0|u)^i_1|vac\rangle^{\l_0-N\hat{1}}_{\l_0}=0,
~~i=2,\cdots,n,\\
&&S(\l_0-N\hat{1},\l_0|u)^i_j|vac\rangle^{\l_0-N\hat{1}}_{\l_0}
=\d^i_j\prod_{k=1}^NR(u-z_k,\l_0-(N-k)\hat{1})^{1j}_{1j}
|vac\rangle^{\l_0-N\hat{1}-\hat{j}}_{\l_0-\hat{j}},\no\\
&&~~~~~~~~~~~~~~~~i,j=2,\cdots,n, \label{Action-2}\eea from
equations (\ref{T-element}) and (\ref{S-element}). Noting that the
diagonal form of $\K(\l_0|u;\xi)$ (\ref{Diag-F}) and the above
equations, we can derive \bea
&&\T(\l_0-N\hat{1},\l_0|u;\xi)^1_1|vac\rangle^{\l_0-N\hat{1}}_{\l_0}
=k(\l_0|u;\xi)_1|vac\rangle^{\l_0-N\hat{1}}_{\l_0},\label{TF-V1}\\
&&\T(\l_0-N\hat{1},\l_0|u;\xi)^i_1|vac\rangle^{\l_0-N\hat{1}}_{\l_0}=0,
~~i=2,\cdots,n.\label{TF-V2} \eea Moreover, after a tedious
calculation, we have (for details, see Appendix C)\bea
&&\T(\l_0-N\hat{1},\l_0|u;\xi)^i_j|vac\rangle^{\l_0-N\hat{1}}_{\l_0}
=\d^i_j\mbox{{\Huge \{}}k(\l_0|u;\xi)_1(R(2u,\l_0-(N-1)\hat{1})^{j1}_{1j}\no\\
&&~~~~~~~~-R(2u,\l_0+\hat{1})^{j1}_{1j}\prod_{k=1}^N
R(u-z_k,\l_0-(N-k)\hat{1})^{1j}_{1j}
R(u+z_k,\l_0-(N-k)\hat{1})^{j1}_{j1})\no\\
&&~~~~~~~~\lt.+k(\l_0|u;\xi)_j \prod_{k=1}^N
R(u-z_k,\l_0-(N-k)\hat{1})^{1j}_{1j}
R(u+z_k,\l_0-(N-k)\hat{1})^{j1}_{j1}\,\rt\}\no\\
&&~~~~~~~~~~\times|vac\rangle^{\l_0-N\hat{1}}_{\l_0},
~~i,j=2,\cdots,n.\label{TF-V3} \eea

Keeping the definition of operators $\A$ (\ref{Def-AB}) and
$\D^i_j$ (\ref{Def-D}) in mind, and using the relations
(\ref{TF-V1})-(\ref{TF-V3}),   we find that the pseudo-vacuum
state given by (\ref{Vac}) satisfies the following equations as
required \bea &&\A(\l_0-N\hat{1}|u)|\O\rangle=k(\l_0|u;\xi)_1
|\O\rangle,\label{A}\\
&&\D^i_j(\l_0-N\hat{1}|u)|\O\rangle=\d^i_j
\b^{(1)}(u)k(\l_0|u+\frac{w}{2};\xi-\frac{w}{2})_j\no\\
&&~~~~~~~~~~~~\times\lt\{\prod_{k=1}^N
\frac{\s(u+z_k)\s(u-z_k)}{\s(u+z_k+w)\s(u-z_k+w)}\rt\}
|\O\rangle,~i,j=2,\cdots,n,\label{D}\\
&&\C_i(\l_0-N\hat{1}|u)|\O\rangle=0,~i=2,\cdots,n,\\
&&\B_i(\l_0-N\hat{1}|u)|\O\rangle\neq 0,~i=2,\cdots,n. \eea The
function $\b^{(1)}(u)$ is \bea \b^{(1)}(u)=
\frac{\s(2u)\s((\l_0)_1w+u+w+\xi)}{\s(2u+w)\s((\l_0)_1w+u+\xi)}.
\eea In  deriving the equation (\ref{D}), we have used the
following equation \bea
k(\l_0|u;\xi)_j-k(\l_0|u;\xi)_1R(2u,\l_0+\hat{1})^{j1}_{1j}=
\b^{(1)}(u)
\frac{\s((\l_0)_{1j}w)\s((\l_0)_jw+\xi-w-u)}{\s((\l_0)_{1j}w+w)
\s((\l_0)_jw+\xi+u)}, \eea which is a consequence of the identity
of (\ref{identity}), the definitions (\ref{Elements2}) and
(\ref{k-def}).

Therefore, we have constructed the pseudo-vacuum state
$|\O\rangle$ which is the common eigenstate of the operators $\A$,
$\D^{i}_{i},\,i=2,\cdots,n,$ and annihilated  by the operators
$\C_i,\,i=2,\cdots,n.$ The operators $\B_i,\,i=2,\cdots,n,$ will
play the role of creation operators to generate the Bethe ansatz
states.
\subsection{Nested Bethe ansatz}
After deriving the relevant commutation relations
(\ref{Rel-1})-(\ref{Rel-3}) and constructing the pseudo-vacuum
state (\ref{Vac}), we now apply the algebraic Bethe ansatz method
(in this case, it is usually called  nested Bethe ansatz) to solve
the eigenvalue problem for the transfer matrices (\ref{trans}) of
$\Zb_n$ Belavin model with open boundary condition specified by
the K-matrices $K^{\pm}(u)$ given in (\ref{K-matrix}),
(\ref{K-matrix1}) and (\ref{boundary-res}). We assume that
$N=n\times l$ with $l$ being a positive integer so that the
algebraic Bethe ansatz method can be applied as in elliptic
integrable models \cite{Bax82,Hou89,Fan96,Hou03}.

Let us introduce a set of integers: \bea N_i=(n-i)\times
l,~~i=0,1,\cdots,n-1,\label{Integer} \eea and $\frac{n(n-1)}{2}l$
complex parameters
$\{\{v^{(i)}_k|~k=1,2,\cdots,N_{i+1}\},~i=0,1,\cdots,n-2\}$ for
convenience. Like usual nested Bethe ansatz method
\cite{Bab82,Sch83,Veg94,Hou03}, the parameters $\{\{v^{(i)}_k\}\}$
will be used to specify the eigenvectors of the corresponding
reduced transfer matrices (see below).  They will be constrained
later by the Bethe ansatz equations (\ref{BA1}) and (\ref{BA2}).
For convenience, we adopt the following convention: \bea
v_k=v^{(0)}_k,~k=1,2,\cdots, N_1. \eea We will seek the common
eigenvectors of the transfer matrices in the form \bea
&&|v_1,\cdots,v_{N_1}\rangle =\sum_{i_1,\cdots,i_{N_1}=2}^n
F^{i_1,i_2,\cdots,i_{N_1}}
\B_{i_1}(\l_0+\hat{i}_1-\hat{1}|v_{1})\B_{i_2}
(\l_0+\hat{i}_1+\hat{i}_2-2\hat{1}|v_2)\no\\
&&~~~~~~~~~~~~~~~~~~~~~~~~~~~~~~\times\cdots\B_{i_{N_1}}
(\l_0+\sum_{k=1}^{N_1}\hat{i}_k-N_1\hat{1}|v_{N_1})
|\O\rangle,\label{Eigenstate} \eea in which
$F^{i_1,i_2,\cdots,i_{N_1}}$ are coefficients to be determined
later by (\ref{Reduced-prob}). The indices in the above equation
should satisfy the following condition: the number of $i_k=j$,
denoted by $\#(j)$, being \bea
  \#(j)=l, ~~~j=2, \cdots, n.\label{Restriction}
\eea With the above restriction, one can derive  \bea
\l_0+\sum_{k=1}^{N_1}\hat{i}_k-N_1\hat{1}&=&
\l_0+l\sum_{k=2}^{n}\hat{k}-(n-1)l\hat{1}\no\\
&=&\l_0+l\sum_{k=1}^{n}\hat{k}-nl\hat{1}=\l_0-N\hat{1},
\label{Restriction1}\eea using the identity (\ref{Vectors}).

Noting the diagonal form of the K-matrix $\tilde{\K}(\l_0|u)$
(\ref{Diag-F}) and its expression (\ref{k-def1}), and the
decomposition form (\ref{De1}) of the transfer matrices, we can
further rewrite the transfer matrices in the following desired
form \bea
&&t(u;\xi)=\sum_{\nu=1}^n\tilde{k}(\l_0|u)_{\nu}\T(\l_0|u;\xi)^{\nu}_{\nu}
\no\\
&&~~~~~
=\tilde{k}(\l_0|u)_1\A(\l_0|u)+\sum_{i=2}^n\tilde{k}(\l_0|u)_i
\T(\l_0|u;\xi)^i_i \no\\
&&~~~~~=\tilde{k}(\l_0|u)_1\A(\l_0|u)+
\sum_{i=2}^n\tilde{k}(\l_0|u)_i
R(2u,\l_0+\hat{1})^{i1}_{1i}\A(\l_0|u)\no\\
&&~~~~~~~~+ \sum_{i=2}^n\tilde{k}(\l_0|u)_i (\T(\l_0|u;\xi)^i_i-
R(2u,\l_0+\hat{1})^{i1}_{1i}\A(\l_0|u))\no\\
&&~~~~~= \sum_{i=1}^n\tilde{k}(\l_0|u)_i
R(2u,\l_0+\hat{1})^{i1}_{1i}\A(\l_0|u)\no\\
&&~~~~~~~~+ \sum_{i=2}^n\tilde{k}^{(1)}(\l_0|u+\frac{w}{2})_i
~\frac{\s((\l_0)_{i1}w-w)}{\s((\l_0)_{i1}w)}(\T(\l_0|u;\xi)^i_i-
R(2u,\l_0+\hat{1})^{i1}_{1i}\A(\l_0|u))\no\\
&&~~~~~=\a^{(1)}(u)\A(\l_0|u)+\sum_{i=2}^n
\tilde{k}^{(1)}(\l_0|u+\frac{w}{2})_i\D(\l_0|u)^i_i.
\label{trans1}\eea We have used the definition (\ref{Def-D}) and
introduced  function $\a^{(1)}(u)$ given by \bea
\a^{(1)}(u)=\sum_{i=1}^n\tilde{k}(\l_0|u)_i
R(2u,\l_0+\hat{1})^{i1}_{1i},\label{function-a}\eea and reduced
K-matrix $\tilde{\K}^{(1)}(\l|u)$ as follows \bea
&&\tilde{\K}^{(1)}(\l_0|u)^j_i=
\d^j_i\tilde{k}^{(1)}(\l_0|u)_i,~~i,j=2,\cdots,n,\label{Reduced-K1}\\
&& \tilde{k}^{(1)}(\l_0|u)_i= \lt\{\prod_{k\neq
i,k=2}^n\frac{\s((\l_0)_{ik}w-w)}{\s((\l_0)_{ik}w)}\rt\}
\frac{\s((\l_0)_iw+\bar{\xi}+u+
\frac{(n-1)w}{2})}{\s((\l_0)_iw+\bar{\xi}-u-\frac{(n-1)w}{2})}.
\label{Reduced-K2} \eea

Now, let us evaluate the action of $\D^i_i(\l_0|u)$ on the state
$|v_1,\cdots,v_{N_1}\rangle$ given in (\ref{Eigenstate}). Many
terms will appear when we move $\D^i_i(\l_0|u)$ from the left to
the right of $\B_i(v_k)$'s. They can be classified into two types:
{\it wanted terms\/} and {\it unwanted terms\/}. The {\it wanted
terms\/} in $\D^i_i(\l_0|u)|v_1,\cdots,v_{N_1}\rangle$ can be
obtained by retaining the first term in the commutation relation
(\ref{Rel-2}). The {\it unwanted terms\/} arising from the second
and third terms of (\ref{Rel-2}), have some $\B(v_k)$ replaced by
$\B(u)$. One {\it unwanted term\/} where $\B(v_1)$ is replaced by
$\B(u)$ can be obtained by using firstly the second and third
terms of (\ref{Rel-2}), then repeatedly using the first term of
(\ref{Rel-1})  and (\ref{Rel-2}). Thanks to the commutation
relation (\ref{Rel-3}), one can easily obtain the other unwanted
terms where the other $\B(v_k)$ is replaced by $\B(u)$. Keeping
the equation (\ref{Restriction1}) and the properties of
pseudo-vacuum state (\ref{A})-(\ref{D}) in mind, we find the
action of $\D^i_i(\l_0|u)$ on the state
$|v_1,\cdots,v_{N_1}\rangle$ \bea
&&\D^i_i(\l_0|u)|v_1,\cdots,v_{N_1}\rangle=
\prod_{k=1}^{N_1}\frac{\s(u-v_k+w)\s(u+v_k+2w)}{\s(u-v_k)\s(u+v_k+w)}
\prod_{k=1}^{N}\frac{\s(u+z_k)\s(u-z_k)}{\s(u+z_k+w)\s(u-z_k+w)}
\no\\
&&~~~~~~~~~~~~~~\times
\b^{(1)}(u){\T^{(1)}}^i_i(\l_0|u+\frac{w}{2};\xi-\frac{w}{2})
^{i_1''\cdots i_{N_1}''}_{i_1\cdots
i_{N_1}}\B_{i_1''}(\l_0+\hat{i}_1''-\hat{1}|v_1)\no\\
&&~~~~~~~~~~~~~~\times
\B_{i_2''}(\l_0+\hat{i}_1''+\hat{i}_2''-2\hat{1}|v_2)
\cdots\B_{i_{N_1}''}(\l_0-N\hat{1}|v_{N_1}) |\O\rangle
F^{i_1,\cdots,i_{N_1}}\no\\
&&~~~~~-\frac{\s(w)\s(2u+2w)\s(u-v_1+(\l_0)_{1\a}w-w)}
{\s(u-v_1)\s(2u+w)\s((\l_0)_{1\a}w-w)} \prod_{k=2}^{N_1}
\frac{\s(v_1-v_k+w)\s(v_1+v_k+2w)}{\s(v_1-v_k)\s(v_1+v_k+w)}\no\\
&&~~~~~~~~~~~~~~\times
\prod_{k=1}^{N}\frac{\s(v_1+z_k)\s(v_1-z_k)}
{\s(v_1+z_k+w)\s(v_1-z_k+w)}
\b^{(1)}(v_1)R(2u+w,\l_0-\hat{i})^{i\a}_{\a i}\no\\
&&~~~~~~~~~~~~~~\times\d^{i_1''}_{\a}
{\T^{(1)}}^{\a}_{i_1}(\l_0+\hat{i}_1-\hat{1}|v_1+\frac{w}{2};\xi-\frac{w}{2})
^{i_2''\cdots i_{N_1}''}_{i_2\cdots i_{N_1}}
\B_{i_1''}(\l_0+\hat{i}_1''-\hat{1}|u)\no\\
&&~~~~~~~~~~~~~~\times
\B_{i_2''}(\l_0+\hat{i}_1''+\hat{i}_2''-2\hat{1}|v_2)
\cdots\B_{i_{N_1}''}(\l_0-N\hat{1}|v_{N_1})|\O\rangle
F^{i_1,\cdots,i_{N_1}}\no\\
&&~~~~~+\frac{\s(w)\s(2v_1)\s(2u+2w)\s(u+v_1+(\l_0)_{1i_1}w)}
{\s(u+v_1+w)\s(2v_1+w)\s(2u+w)\s((\l_0)_{1i_1}w-w)}
k(\l_0|v_1;\xi)_1\no\\
&&~~~~~~~~~~~~~~\times \prod_{k=2}^{N_1}
\frac{\s(v_1+v_k)\s(v_1-v_k+w)} {\s(v_1+v_k+w)\s(v_1-v_k)}
R(2u+w,\l_0-\hat{i})^{i\,\,\,\,i_1}_{i_1\,i}\B_{i_1}(\l_0+\hat{i}_1-\hat{1}|u)
\no\\
&&~~~~~~~~~~~~~~\times
\B_{i_2}(\l_0+\hat{i}_1+\hat{i}_2-2\hat{1}|v_2)\cdots
\B_{i_{N_1}}(\l_0-N\hat{1}|v_{N_1})|\O\rangle
F^{i_1,\cdots,i_{N_1}}\no\\
&&~~~~~+o.u.t., \label{Action-D}\eea where {\it o.u.t\/} stands
for the {\it other unwanted terms\/}. We adopt here and in the
following part of this subsection the convention that the repeated
indices imply summation over $2,\cdots,n$ (but we do not take the
summation for the index $i$ in the above equation). We have
introduced reduced monodromy matrix $\T^{(1)}(\l|u;\xi)$: \bea
&&{\T^{(1)}}^{j}_{i}
(\l|u+\frac{w}{2};\xi-\frac{w}{2})^{i_1''\cdots
i_{N_1}''}_{i_1\cdots
i_{N_1}}=R(u+\frac{w}{2}+z^{(1)}_1,\l-\hat{i})^{j\,\,\,\,\,\,i''_1}_{\b_{N_1-1}\,i'_1}\no\\
&&~~~~~~~~~~\times
R(u+\frac{w}{2}+z^{(1)}_2,\l-\hat{i}+\hat{i'}_1)^{\b_{N_1-1}\,i''_2}_{\b_{N_1-2}\,i'_2}
 \cdots R(u+\frac{w}{2}+z^{(1)}_{N_1},\l-\hat{i}+\sum_{k=1}^{N_1-1}
\hat{i'}_k)^{\b_1\,i''_{N_1}}_{\nu\,\,i'_{N_1}}\no\\
&&~~~~~~~~~~\times k(\l|u+\frac{w}{2};\xi-\frac{w}{2})_{\nu}
R(u+\frac{w}{2}-z^{(1)}_{N_1},\l+\sum_{k=1}^{N_1}
\hat{i}_k)^{i'_{N_1}\,\nu}_{i_{N_1}\,\a_{N_1-1}}\no\\
&&~~~~~~~~~~\times
R(u+\frac{w}{2}-z^{(1)}_{N_1-1},\l+\sum_{k=1}^{N_1-1}
\hat{i}_k)^{i'_{N_1-1}\,\a_{N_1-1}}_{i_{N_1-1}\,\a_{N_1-2}}\cdots
R(u+\frac{w}{2}-z^{(1)}_1,\l+\hat{i}_1)^{i'_1\,\a_1}_{i_1\,i},\no\\
\label{Reduce-mon} \eea where \bea
z^{(1)}_k=v_k+\frac{w}{2},~~k=1,\cdots,N_1.\eea The following
property of R-matrix has been used to derive the equation
(\ref{Action-D}) \bea R(u,\l+\hat{1})^{ij}_{kl}=R(u,\l)^{ij}_{kl},
~~~i,j,k,l>1.\no\eea

Similarly, we find the action of $\A(\l_0|u)$ on the state
$|v_1,\cdots,v_{N_1}\rangle$ \bea
&&\A(\l_0|u)|v_1,\cdots,v_{N_1}\rangle=k(\l_0|u;\xi)_1
\prod_{k=1}^{N_1}\frac{\s(u+v_k)\s(u-v_k-w)}{\s(u+v_k+w)\s(u-v_k)}
|v_1,\cdots,v_{N_1}\rangle\no\\
&&~~~~~-\frac{\s(w)\s(2v_1)\s(u-v_1-(\l_0)_{1i_1}w+w)}
{\s(u-v_1)\s(2v_1+w)\s((\l_0)_{1i_1}w-w)} k(\l_0|u;\xi)_1
\prod_{k=2}^{N_1}
\frac{\s(v_1+v_k)\s(v_1-v_k-w)}{\s(v_1+v_k+w)\s(v_1-v_k)}\no\\
&&~~~~~~~~~~~~~~\times\B_{i_1}(\l_0+\hat{i}_1-\hat{1}|u)
\B_{i_2}(\l_0+\hat{i}_1+\hat{i}_2-2\hat{1}|v_2)\cdots\no\\
&&~~~~~~~~~~~~~~\times\B_{i_{N_1}}(\l_0-N\hat{1}|v_{N_1})|\O\rangle
F^{i_1,\cdots,i_{N_1}}\no\\
&&~~~~~-\frac{\s(w)\s(u+v_1+(\l_0)_{\a 1}w+2w)}
{\s(u+v_1+w)\s((\l_0)_{\a 1}w+w)} \b^{(1)}(v_1) \prod_{k=2}^{N_1}
\frac{\s(v_1-v_k+w)\s(v_1+v_k+2w)}{\s(v_1-v_k)\s(v_1+v_k+w)}\no\\
&&~~~~~~~~~~~~~~\times
\prod_{k=1}^{N}\frac{\s(v_1+z_k)\s(v_1-z_k)}
{\s(v_1+z_k+w)\s(v_1-z_k+w)}
\d^{{i}_1''}_{\a}{\T^{(1)}}^{\a}_{i_1}
(\l_0+\hat{i}_1-\hat{1}|v_1+\frac{w}{2};\xi-\frac{w}{2})^{i_2''\cdots
i_{N_1}''}_{i_2\cdots i_{N_1}}\no\\
&&~~~~~~~~~~~~~~\times \B_{i_1''}(\l_0+\hat{i}_1''-\hat{1}|u)
\B_{i_2''}(\l_0+\hat{i}_1''+\hat{i}_2''-2\hat{1}|v_2)\no\\
&&~~~~~~~~~~~~~~\times
\cdots\B_{i_{N_1}''}(\l_0-N\hat{1}|v_{N_1})|\O\rangle
F^{i_1,\cdots,i_{N_1}}\no\\
&&~~~~~+o.u.t..\label{Action-A}\eea Using the results when
$\A(\l_0|u)$ and $\D^i_i(\l_0|u)$ act on the state
$|v_1,\cdots,v_{N_1}\rangle$ (\ref{Action-A}) and
(\ref{Action-D}), and the equation (\ref{trans1}), we can find the
action of the transfer matrices on the state
$|v_1,\cdots,v_{N_1}\rangle$ \bea
&&t(u;\xi)|v_1,\cdots,v_{N_1}\rangle
=(\a^{(1)}(u)\A(\l_0|u)+\sum_{i=2}^n
\tilde{k}^{(1)}(\l_0|u+\frac{w}{2})_i\D(\l_0|u)^i_i)
|v_1,\cdots,v_{N_1}\rangle\no\\
&&~~~~~=\a^{(1)}(u)k(\l_0|u;\xi)_1
\prod_{k=1}^{N_1}\frac{\s(u+v_k)\s(u-v_k-w)}{\s(u+v_k+w)\s(u-v_k)}
|v_1,\cdots,v_{N_1}\rangle\no\\
&&~~~~~~~~~~~~+ \b^{(1)}(u)
\prod_{k=1}^{N_1}\frac{\s(u-v_k+w)\s(u+v_k+2w)}{\s(u-v_k)\s(u+v_k+w)}
\prod_{k=1}^{N}\frac{\s(u+z_k)\s(u-z_k)}
{\s(u+z_k+w)\s(u-z_k+w)}\no\\
&&~~~~~~~~~~~~~~~~~~\times t^{(1)}(u+\frac{w}{2};\xi-\frac{w}{2})
^{i_1''\cdots i_{N_1}''}_{i_1\cdots i_{N_1}}|\O\rangle
F^{i_1,\cdots,i_{N_1}}\no\\
&&~~~~~~~~~~~~+u.t., \label{Main}\eea where {\it u.t.} stands for
the {\it unwanted terms\/}. Reduced transfer matrices
$t^{(1)}(u;\xi)$ are given in terms of the reduced monodromy
matrix (\ref{Reduce-mon}) and the reduced K-matrix
(\ref{Reduced-K1}) and (\ref{Reduced-K2}) \bea t^{(1)}(u;\xi)
=\sum_{i=2}^n\tilde{k}^{(1)}(\l_0|u)_i
{\T^{(1)}}^i_i(\l_0|u;\xi).\label{reduce-trans}\eea The equation
(\ref{Main}) tells that the state $ |v_1,\cdots,v_{N_1}\rangle$ is
not an eigenvector of the transfer matrices $t(u;\xi)$ {\em
unless\/} $F's$ are the eigenvectors of the reduced transfer
matrices $t^{(1)}(u;\xi)$ as (\ref{Reduced-prob}). The condition
that the {\it unwanted} terms  should cancel each other, will give
a restriction on the $N_1$ parameters $\lt\{v_k\rt\}$, the
so-called Bethe ansatz equation. Hence we arrive at the final
results: \bea t(u;\xi)|v_1,\cdots,v_{N_1}\rangle=\L(u;\xi,\{v_k\})
|v_1,\cdots,v_{N_1}\rangle.\eea The eigenvalue reads \bea
&&\L(u;\xi,\{v_k\})=\a^{(1)}(u)\frac{\s((\l_0)_1w+\xi-u)}
{\s((\l_0)_1w+\xi+u)}\prod_{k=1}^{N_1}\frac{\s(u+v_k)\s(u-v_k-w)}
{\s(u+v_k+w)\s(u-v_k)}\no\\
&&~~~~~~+\frac{\s(2u)\s((\l_0)_1w+u+w+\xi)}
{\s(2u+w)\s((\l_0)_1w+u+\xi)}
\prod_{k=1}^{N_1}\frac{\s(u-v_k+w)\s(u+v_k+2w)}
{\s(u-v_k)\s(u+v_k+w)}\no\\
&&~~~~~~~~~~~~\times \prod_{k=1}^{N_0}\frac{\s(u+z_k)\s(u-z_k)}
{\s(u+z_k+w)\s(u-z_k+w)}
\L^{(1)}(u+\frac{w}{2};\xi-\frac{w}{2},\{v^{(1)}\}),\label{Eigenvalue1}\eea
in which the functions $\a^{(1)}(u)$ is given in
(\ref{function-a}) and $\L^{(1)}(u;\xi,\{v^{(1)}_k\})$ is the
eigenvalue of the reduced transfer matrices defined in
(\ref{reduce-trans}) \bea t^{(1)}(u;\xi)^{i_1\cdots
i_{N_1}}_{i'_1\cdots i'_{N_1}}F^{i'_1\cdots
i'_{N_1}}=\L^{(1)}(u;\xi,\{v^{(1)}_k\})F^{i_1\cdots
i_{N_1}}.\label{Reduced-prob}\eea The parameters
$\{v_k|k=1,\cdots,N_1\}$ satisfy so-called Bethe ansatz equation
\bea &&\a^{(1)}(v_s)\frac{\s((\l_0)_1w+\xi-v_s)\s(2v_s+w)}
{\s((\l_0)_1w+\xi+v_s+w)\s(2v_s+2w)} \prod_{k\ne s,k=1}^{N_1}
\frac{\s(v_s+v_k)\s(v_s-v_k-w)} {\s(v_s+v_k+2w)\s(v_s-v_k+w)}\no\\
&&~~~~~~= \prod_{k=1}^{N} \frac{\s(v_s+z_k)\s(v_s-z_k)}
{\s(v_s+z_k+w)\s(v_s-z_k+w)}
\L^{(1)}(v_s+\frac{w}{2};\xi-\frac{w}{2},\{v^{(1)}_k\}).
\label{BA1}\eea The parameters $\{v^{(1)}_k|k=1,\cdots,N_2\}$ will
be specified later by the further Bethe ansatz equation
(\ref{BA2}).

The diagonalization of the transfer matrices $t(u;\xi)$ of $\Zb_n$
Belavin model  with open boundary condition specified by the
K-matrices $K^{\pm}(u)$ given in (\ref{K-matrix}),
(\ref{K-matrix1}) and (\ref{boundary-res})
 is now reduced to the diagonalization of the
reduced transfer matrices $t^{(1)}(u;\xi)$ in
(\ref{Reduced-prob}). The explicit expression of $t^{(1)}(u;\xi)$
given in (\ref{reduce-trans}) and (\ref{Reduce-mon}) implies that
$t^{(1)}(u;\xi)$ can be considered as the transfer matrices of a
$\Zb_{n-1}$ Belavin model  with open boundary condition: the
corresponding quantum space (resp. inhomogeneous parameters
$\{z_k\}$ in (\ref{Mon-V}) ) is replaced by $(\Cb^{n-1})^{N_1}$
(resp. $\{z^{(1)}_k\}$); the boundary parameter $\xi$ for
$\K(\l_0|u;\xi)$ given in (\ref{k-def}) is replaced by
$\xi^{(1)}=\xi-\frac{w}{2}$ (see (\ref{Reduce-mon})); at the same
time the corresponding $\tilde{\K}(\l_0|u)$ is replaced by the
reduced one given in (\ref{Reduced-K1}) and (\ref{Reduced-K2}). We
can use the same method to find the eigenvalue of $t^{(1)}(u;\xi)$
as we have done for the diagonalization of $t(u;\xi)$.  Repeating
the above procedure further $n-2$ times, one can reduce to the
last reduced transfer matrices $t^{(n-1)}(u;\xi)$ which are
trivial to get the eigenvalues. This is so-called nested Bethe
ansatz. At the same time we need to introduce the
$\frac{n(n-1)}{2}l$ parameters $\{\{
v_k^{(i)}|~k=1,2,\cdots,N_{i+1}\},~i=0,1,\cdots,n-2\}$ to specify
the eigenvectors of the corresponding reduced transfer matrices
$t^{(i)}(u;\xi)$  (including the original one
$t(u;\xi)=t^{(0)}(u;\xi)$), and the further reduced  K-matrices
$\tilde{\K}^{(m)}(\l_0|u)$ (including the original one
$\tilde{\K}(\l_0|u)=\tilde{\K}^{(0)}(\l_0|u)$) as we have done in
(\ref{Reduced-K1}) and (\ref{Reduced-K2}). The reduced K-matrices
are given  as follows \bea &&\tilde{\K}^{(m)}(\l_0|u)^j_i=
\d^j_i\tilde{k}^{(1)}(\l_0|u)_i,~~i,j=m+1,\cdots,n,~~m=0,\cdots
n-1,
\label{Reduced-K3}\\
&& \tilde{k}^{(m)}(\l_0|u)_i= \lt\{\prod_{k\neq
i,k=m+1}^n\frac{\s((\l_0)_{ik}w-w)}{\s((\l_0)_{ik}w)}\rt\}
\frac{\s((\l_0)_iw+\bar{\xi}+u+
\frac{(n-m)w}{2})}{\s((\l_0)_iw+\bar{\xi}-u-\frac{(n-m)w}{2})}.
\label{Reduced-K4} \eea Like (\ref{function-a}), we introduce a
set of functions $\{\a^{(m)}(u)|m=1,\cdots,n-1\}$ related to the
reduced K-matrices $\tilde{\K}^{(m)}(\l_0|u)$ \bea
\a^{(m)}(u)=\sum_{i=m}^{n}R(2u,\l_0+\hat{m})^{im}_{mi}
\tilde{k}^{(m-1)}(\l_0|u)_i,~~m=1,\cdots,n.\eea Finally, we obtain
all the eigenvalues of the reduced transfer matrices
$t^{(i)}(u;\xi)$ with the eigenvalue
$\L^{(i)}(u;\xi,\{v^{(i)}_{k}\})$ in a recurrence form \bea
&&\L^{(i)}(u;\xi^{(i)},\{v^{(i)}_k\})=\a^{(i+1)}(u)
\frac{\s((\l_0)_{i+1}w+\xi^{(i)}-u)}
{\s((\l_0)_{i+1}w+\xi^{(i)}+u)}
\prod_{k=1}^{N_{i+1}}\frac{\s(u+v^{(i)}_k)\s(u-v^{(i)}_k-w)}
{\s(u+v^{(i)}_k+w)\s(u-v^{(i)}_k)}\no\\
&&~~~~~~+\frac{\s(2u)\s((\l_0)_{i+1}w+u+w+\xi^{(i)})}
{\s(2u+w)\s((\l_0)_{i+1}w+u+\xi^{(i)})}
\prod_{k=1}^{N_{i+1}}\frac{\s(u-v^{(i)}_k+w)\s(u+v^{(i)}_k+2w)}
{\s(u-v^{(i)}_k)\s(u+v^{(i)}_k+w)}\no\\
&&~~~~~~~~~~~~\times
\prod_{k=1}^{N_i}\frac{\s(u+z^{(i)}_k)\s(u-z^{(i)}_k)}
{\s(u+z^{(i)}_k+w)\s(u-z^{(i)}_k+w)}
\L^{(i+1)}(u+\frac{w}{2};\xi^{(i)}-\frac{w}{2},\{v^{(i+1)}\}),\no\\
&&~~~~~~~~~~~~i=1,\cdots,n-2, \label{Eigenvalue2}\\
&&\L^{(n-1)}(u;\xi^{(n-1)})=\frac{\s((\l_0)_nw+\bar{\xi}+u+\frac{w}{2})
\s((\l_0)_nw+\xi^{(n-1)}-u)}
{\s((\l_0)_nw+\bar{\xi}-u-\frac{w}{2})
\s((\l_0)_nw+\xi^{(n-1)}+u)}.\label{Eigenvalue3} \eea The reduced
boundary parameters $\{\xi^{(i)}\}$ and inhomogeneous parameters
$\{z^{(i)}_k\}$ are given by \bea
\xi^{(i+1)}=\xi^{(i)}-\frac{w}{2},~~z^{(i+1)}_k=v^{(i)}_k+\frac{w}{2},
~~i=0,\cdots,n-2.\eea We adopt the convention: $\xi=\xi^{(0)}$,
$z^{(0)}_k=z_k$. The $\{\{v^{(i)}_k\}\}$ satisfy the following
Bethe ansatz equations \bea
&&\a^{(i+1)}(v^{(i)}_s)\frac{\s((\l_0)_{i+1}w+\xi^{(i)}-v^{(i)}_s)
\s(2v^{(i)}_s+w)}
{\s((\l_0)_{i+1}w+\xi^{(i)}+v^{(i)}_s+w)\s(2v^{(i)}_s+2w)}\no\\
&&~~~~~~~~~~~~~~~~\times\prod_{k\ne s,k=1}^{N_{i+1}}
\frac{\s(v^{(i)}_s+v^{(i)}_k)\s(v^{(i)}_s-v^{(i)}_k-w)}
{\s(v^{(i)}_s+v^{(i)}_k+2w)\s(v^{(i)}_s-v^{(i)}_k+w)}\no\\
&&~~~~~~= \prod_{k=1}^{N_i}
\frac{\s(v^{(i)}_s+z^{(i)}_k)\s(v^{(i)}_s-z^{(i)}_k)}
{\s(v^{(i)}_s+z^{(i)}_k+w)\s(v^{(i)}_s-z^{(i)}_k+w)}
\L^{(i+1)}(v^{(i)}_s+\frac{w}{2};\xi^{(i)}-\frac{w}{2},\{v^{(i+1)}_k\})
,\no\\
&&~~~~~~~~~~~~~~~~~~i=1,\cdots,n-2.\label{BA2}\eea

\section{Conclusions}
\label{Con} \setcounter{equation}{0}

We have studied $\Zb_n$ Belavin model with integrable open
boundary condition which are described by the boundary K-matrix
$K^-(u)$ given in (\ref{K-matrix}) and its dual $K^+(u)$ given in
(\ref{K-matrix1}) with restriction (\ref{boundary-res}). The total
number of the {\it independent\/} free boundary parameters among
$\xi,\,\bar\xi$ and $(\l_0)_i,\,i=1,\cdots,n$, is actually $n+1$
\footnote{One can fix one of $\{(\l_0)_i\}$, for an example,
$(\l_0)_n=0$ by shifting $\xi\longrightarrow \xi-(\l_0)_nw$ in
(\ref{k-def}).}. Although the K-matrices are non-diagonal in the
vertex form, they {\it simultaneously\/} become diagonal ones in
the face type after the face-vertex transformation which are given
in (\ref{Diag-F})-(\ref{k-def1}) (c.f. (\ref{K-matrix}) and
(\ref{K-matrix1})).  This  fact {\it enables\/} us to successfully
construct the corresponding pseudo-vacuum state $|\O\rangle$
(\ref{Vac}) and apply the algebraic Bethe ansatz method to
diagonalize the corresponding {\it double-row transfer matrices\/}
after using the intertwiner vectors and face-vertex correspondence
relation. The eigenvalues of the transfer matrices and associated
Bethe ansatz equations are given by (\ref{Eigenvalue1}),
(\ref{Eigenvalue2})-(\ref{Eigenvalue3}), and (\ref{BA1}),
(\ref{BA2}). For the special case of $n=2$ (or the eight-vertex
model case), our result recovers that of \cite{Fan96}.

\section*{Acknowledgements}
We thank H. Fan for useful discussion. W.\,-L. Yang is supported
by the Japan Society for the Promotion of Science.

\section*{Appendix A: The exchange relation of
$\T$} \setcounter{equation}{0}
\renewcommand{\theequation}{A.\arabic{equation}}

The starting point for deriving the exchange relations
(\ref{RE-F}) among $\T(\l|u;\xi)^{\nu}_{\mu}$ is the exchange
relation (\ref{Relation-Re}). Multiplying  both sides of equation
(\ref{Relation-Re}) from the right by
$\phi_{\l+\hat{i}_3,\,\l}(-u_1) \otimes
\phi_{\l+\hat{i}_3+\hat{j}_3,\,\l+\hat{i}_3}(-u_2)$, we can derive
the following equations with the help of the face-vertex
correspondence relation (\ref{Face-vertex}), the definition of
$\T(\l|u;\xi)^{\nu}_{\mu}$ (\ref{Mon-F}), and relation
(\ref{Int4})\bea &&L.H.S.
=R^B_{12}(u_1-u_2)\mathbb{T}_1(u_1)R^B_{21}(u_1+u_2)
(\phi_{\l+\hat{i}_3,\,\l}(-u_1)\otimes \mathbb{T}(u_2)
\phi_{\l+\hat{i}_3+\hat{j}_3,\,\l+\hat{i}_3}(-u_2))\no\\
&&~~~~~~=R^B_{12}(u_1-u_2)\mathbb{T}_1(u_1)R^B_{21}(u_1+u_2)
(\phi_{\l+\hat{i}_3,\,\l}(-u_1)\otimes 1)\no\\
&&~~~~~~~~~~~~~~~~~~~\times(1\otimes \{\sum_{j_2}
\phi_{\l+\hat{i}_3+\hat{j}_2,\,\l+\hat{i}_3}(u_2)
\tilde{\phi}_{\l+\hat{i}_3+\hat{j}_2,\,\l+\hat{i}_3}(u_2)
\mathbb{T}(u_2)
\phi_{\l+\hat{i}_3+\hat{j}_3,\,\l+\hat{i}_3}(-u_2)\})\no\\
&&~~~~~~=\sum_{j_2}R^B_{12}(u_1-u_2)\mathbb{T}_1(u_1)R^B_{21}(u_1+u_2)
(\phi_{\l+\hat{i}_3,\,\l}(-u_1)\otimes
\phi_{\l+\hat{i}_3+\hat{j}_2,\,\l+\hat{i}_3}(u_2))\no\\
&&~~~~~~~~~~~~~~~~~~~\times\T(\l+\hat{i}_3+\hat{j}_3|u_2;\xi)^{j_2}_{j_3}
\no\\
&&~~~~~~=\sum_{i_2}\sum_{j_1,j_2}R^B_{12}(u_1-u_2)\mathbb{T}_1(u_1)
R(u_1+u_2,\l)^{j_1\,i_2}_{j_2\,i_3}
(\phi_{\l+\hat{i}_3+\hat{j}_2,\,\l+\hat{j}_1}(-u_1)\otimes
\phi_{\l+\hat{j}_1,\,\l}(u_2))\no\\
&&~~~~~~~~~~~~~~~~~~\times
\T(\l+\hat{i}_3+\hat{j}_3|u_2;\xi)^{j_2}_{j_3}
\no\\
&&~~~~~~~~~~~~~\vdots\no\\
&&~~~~~~~~~~~~~\vdots\no\\
&&~~~~~~=\sum_{i_0,j_0} (\phi_{\l+\hat{i}_0,\,\l}(u_1)\otimes
\phi_{\l+\hat{i}_0+\hat{j}_0,\,\l+\hat{i}_0}(u_2))\sum_{i_1,i_2}
\sum_{j_1,j_2}R(u_1-u_2,\l)^{i_0\,j_0}_{i_1\,j_1}
\T(\l+\hat{i}_2+\hat{j}_1|u_1;\xi)^{i_1}_{i_2}\no\\
&&~~~~~~~~~~~~~~~~~~\times R(u_1+u_2,\l)^{j_1\,i_2}_{j_2\,i_3}
\T(\l+\hat{i}_3+\hat{j}_3|u_2;\xi)^{j_2}_{j_3}.\label{LHS} \eea
Similarly, we have \bea &&R.H.S.= \sum_{i_0,j_0}
(\phi_{\l+\hat{i}_0,\,\l}(u_1)\otimes
\phi_{\l+\hat{i}_0+\hat{j}_0,\,\l+\hat{i}_0}(u_2))\sum_{i_1,i_2}
\sum_{j_1,j_2}\T(\l+\hat{i}_0+\hat{j}_1|u_2;\xi)^{j_0}_{j_1}
R(u_1+u_2,\l)^{i_0\,j_1}_{i_1\,j_2}
\no\\
&&~~~~~~~~~~~~~~~~~~\times
\T(\l+\hat{i}_2+\hat{j}_2|u_1;\xi)^{i_1}_{i_2}
R(u_1-u_2,\l)^{j_2\,i_2}_{j_3\,i_3}.\label{RHS} \eea Noting the
fact that intertwiners are linearly independent of each other with
a generic $\l$ \cite{Jim87}, we obtain the exchange relation
(\ref{RE-F}) by comparing (\ref{LHS}) with (\ref{RHS}).

\section*{Appendix B: The relevant commutation relations}
\setcounter{equation}{0}
\renewcommand{\theequation}{B.\arabic{equation}}

Let us adopt the following notation for convenience \bea
&&A(\l|u)=\T(\l|u;\xi)^1_1,~~B_i(\l|u)=\T(\l|u;\xi)^1_i,~i=2,\cdots,n,\\
&&D^j_i(\l|u)=\T(\l|u;\xi)^j_i,~i,j=2,\cdots,n.\eea The starting
point for deriving the commutation relations among
$\A(u),~\D^j_i(u)$ and $\B_i(u)$ $(i,j=2,\cdots,n)$ is the
exchange relation (\ref{RE-F}).

For $i_0=j_0=j_3=1,~ i_3=i\neq 1$, we obtain \bea
&&A(\l+2\hat{1}|v)B_i(\l+\hat{1}+\hat{i}|u)=
\frac{\s(u+v)\s(u-v+w)}{\s(u+v+w)\s(u-v)}
B_i(\l+\hat{1}+\hat{i}|u)A(\l+\hat{1}+\hat{i}|v)\no\\
&&~~~~~~~~~~~~~~~~~~~~~~-\frac{\s(w)\s(u+v)\s(\l_{1i}w+u-v)}
{\s(u-v)\s(u+v+w)\s(\l_{1i}w)}
B_i(\l+\hat{1}+\hat{i}|v)A(\l+\hat{1}+\hat{i}|u)\no\\
&&~~~~~~~~~~~~~~~~~~~~~~-\sum_{j=2}^{n}\frac{\s(w)\s(u+v+\l_{j1}w)}
{\s(u+v+w)\s(\l_{j1}w)}
B_j(\l+\hat{1}+\hat{j}|v)D^j_i(\l+\hat{1}+\hat{i}|u).\label{AB}
\eea One can derive  the following equation \bea
&&\frac{R(u+v,\l)^{1\,i}_{1\,i}R(u-v,\l)^{i\,1}_{1\,i}}
{R(u+v,\l)^{1\,1}_{1\,1}R(u-v,\l)^{1\,i}_{1\,i}} +
\frac{R(u+v,\l)^{1\,i}_{i\,1}R(2u,\l+\hat{1})^{i\,1}_{1\,i}}
{R(u+v,\l)^{1\,1}_{1\,1}R(2u,\l+\hat{1})^{1\,1}_{1\,1}}\no\\
&&~~~~~~~~~~~~~~~~=\frac{\s(w)\s(2u)\s(\l_{1i}w+u-v+w)}
{\s(u-v)\s(2u+w)\s(\l_{1i}w+w)},\label{B-4}\eea from the identity
(\ref{identity}), the definitions (\ref{Elements1}) and
(\ref{Elements2}). The commutation relation (\ref{Rel-1}) is a
simple consequence of the equations (\ref{AB}) and (\ref{B-4}),
the definitions (\ref{Def-AB}) and (\ref{Def-D}).

For $i_0=k\neq 1$, $j_0=1$, $i_3=i\neq 1$ and $j_3=j\neq 1$, we
obtain \bea
&&D^k_i(\l+\hat{1}+\hat{i}|u)B_j(\l+\hat{j}+\hat{i}|v)=
\sum_{\a_1,\a_2,\b_1,\b_2=2}^n
\frac{R(u+v,\l)^{k\,\,\,\b_2}_{\a_2\,\b_1}R(u-v,\l)^{\b_1\,\a_1}_{j\,\,\,\,i}}
{R(u-v,\l)^{k\,1}_{k\,1}R(u+v,\l)^{1\,i}_{1\,i}}\no\\
&&~~~~~~~~~~~~~~~~~~~~~~~~~~~~~~~~~~~~~~~~~~~~~~~~~~~~~~~\times
B_{\b_2}(\l+\hat{k}+\hat{\b}_2|v)
D^{\a_2}_{\a_1}(\l+\hat{i}+\hat{j}|u)\no\\
&&~~~~~~~~~~~~~~~~~~~~+\sum_{\a=2}^n
\frac{R(u+v,\l)^{k\,1}_{1\,k}R(u-v,\l)^{k\,\a}_{j\,\,\,i}}
{R(u-v,\l)^{k\,1}_{k\,1}R(u+v,\l)^{1\,i}_{1\,i}}
A(\l+\hat{k}+\hat{1}|v)
B_{\a}(\l+\hat{i}+\hat{j}|u)\no\\
&&~~~~~~~~~~~~~~~~~~~~-\sum_{\a,\b=2}^n
\frac{R(u-v,\l)^{k\,1}_{1\,k}R(u+v,\l)^{k\,\a}_{\b\,i}}
{R(u-v,\l)^{k\,1}_{k\,1}R(u+v,\l)^{1\,i}_{1\,i}}
B_{\a}(\l+\hat{k}+\hat{\a}|u)
D^{\b}_{j}(\l+\hat{i}+\hat{j}|v)\no\\
&&~~~~~~~~~~~~~~~~~~~~-
\frac{R(u-v,\l)^{k\,1}_{1\,k}R(u+v,\l)^{k\,1}_{1\,i}}
{R(u-v,\l)^{k\,1}_{k\,1}R(u+v,\l)^{1\,i}_{1\,i}}
A(\l+\hat{k}+\hat{1}|u) B_{j}(\l+\hat{i}+\hat{j}|v).
\label{DB}\eea In order to separate the contribution of $A$ and
$D^j_i$ in the above relations, we need to introduce
(cf.\cite{Skl88}): \bea
\bar{D}^k_i(\l|u)=D^k_i(\l|u)-\d^k_iR(2u,\l+\hat{1})^{k\,1}_{1\,k}A(\l|u),
~i,k=2,\cdots,n.\eea We can derive the commutation relations
 among $\bar{D}^k_i$ and $B_j$ from (\ref{DB}) \bea
&&\bar{D}^k_i(\l|u)B_j(\l+\hat{j}-\hat{1}|v)=
\sum_{\a_1,\a_2,\b_1,\b_2=2}^n
\frac{R(u+v,\l-\hat{1}-\hat{i})^{k\,\,\,\b_2}_{\a_2\,\b_1}
R(u-v,\l-\hat{1}-\hat{i})^{\b_1\,\a_1}_{j\,\,\,\,i}}
{R(u-v,\l-\hat{1}-\hat{i})^{k\,1}_{k\,1}
R(u+v,\l-\hat{1}-\hat{i})^{1\,i}_{1\,i}}\no\\
&&~~~~~~~~~~~~~~~~~~~~~~~~~~~~~~~~~~~~~~~~~~~~~~~~~~~~~~~\times
B_{\b_2}(\l+\hat{k}+\hat{\b}_2-\hat{1}-\hat{i}|v)
\bar{D}^{\a_2}_{\a_1}(\l-\hat{1}+\hat{j}|u)\no\\
&&~~~~~~-\sum_{\a,\b=2}^n
\frac{R(u-v,\l-\hat{1}-\hat{i})^{k\,1}_{1\,k}
R(u+v,\l-\hat{1}-\hat{i})^{k\,\a}_{\b\,i}}
{R(u-v,\l-\hat{1}-\hat{i})^{k\,1}_{k\,1}
R(u+v,\l-\hat{1}-\hat{i})^{1\,i}_{1\,i}}
B_{\a}(\l+\hat{\b}-\hat{1}|u)
\bar{D}^{\b}_{j}(\l-\hat{1}+\hat{j}|v)\no\\
&&~~~~~~+\sum_{\a,\b_1,\b_2=2}^n
\frac{R(u+v,\l-\hat{1}-\hat{i})^{k\,\,\b_2}_{\a\,\b_1}
R(u-v,\l-\hat{1}-\hat{i})^{\b_1\,\a}_{j\,\,i}}
{R(u-v,\l-\hat{1}-\hat{i})^{k\,1}_{k\,1}
R(u+v,\l-\hat{1}-\hat{i})^{1\,i}_{1\,i}}
R(2u,\l+\hat{j})^{\a\,1}_{1\,\a}\no\\
&&~~~~~~~~~~~~~~~~~~~~~~~~~~~~~~~~~~~~~~~~~~~~~~~~~~~~~~~\times
B_{\b_2}(\l-\hat{1}-\hat{i}+\hat{\b}_2+\hat{k}|v)
A(\l-\hat{1}+\hat{j}|u)\no\\
&&~~~~~~-\sum_{\a=2}^n
\frac{R(u-v,\l-\hat{1}-\hat{i})^{k\,1}_{1\,k}
R(u+v,\l-\hat{1}-\hat{i})^{k\,\a}_{j\,\,i}}
{R(u-v,\l-\hat{1}-\hat{i})^{k\,1}_{k\,1}
R(u+v,\l-\hat{1}-\hat{i})^{1\,i}_{1\,i}}
R(2v,\l+\hat{j})^{j\,1}_{1\,j}\no\\
&&~~~~~~~~~~~~~~~~~~~~~~~~~~~~~~~~~~~~~~~~~~~~~~~~~~~~~~~\times
B_{\a}(\l-\hat{1}-\hat{i}+\hat{\a}+\hat{k}|u)
A(\l-\hat{1}+\hat{j}|v)\no\\
&&~~~~~~+\sum_{\a=2}^n
\frac{R(u+v,\l-\hat{1}-\hat{i})^{k\,1}_{1\,k}
R(u-v,\l-\hat{1}-\hat{i})^{k\,\a}_{j\,i}}
{R(u-v,\l-\hat{1}-\hat{i})^{k\,1}_{k\,1}
R(u+v,\l-\hat{1}-\hat{i})^{1\,i}_{1\,i}}
A(\l+\hat{k}-\hat{i}|v)B_{\a}(\l+\hat{j}-\hat{1}|u)\no\\
&&~~~~~~-\frac{\s(u-v+w)\s(u+v+w)\s(2u+\l_{1k}w)\s(\l_{1k}w)\s(w)}
{\s(u-v)\s(u+v)\s(2u+w)\s(\l_{1k}w-w)\s(\l_{1k}w+w)}\no\\
&&~~~~~~~~~~~~~~~~~~~~~~~~~~~~~~~~~~~~~~~~~~~~~~~~~~~~~~~\times
\d^k_i
A(\l-\hat{i}+\hat{k}|u)B_j(\l+\hat{j}-\hat{1}|v).\no\\
\eea We have used the following equation \bea
&&\frac{R(u-v,\l-\hat{1}-\hat{i})^{k\,1}_{1\,k}
R(u+v,\l-\hat{1}-\hat{i})^{k\,1}_{1\,k}}
{R(u-v,\l-\hat{1}-\hat{i})^{k\,1}_{k\,1}
R(u+v,\l-\hat{1}-\hat{i})^{1\,i}_{1\,i}}
+R(2u,\l+\hat{1})^{k\,1}_{1\,k}\no\\
&&~~~~~~=\frac{\s(u-v+w)\s(u+v+w)\s(2u+\l_{1k}w)\s(\l_{1k}w)\s(w)}
{\s(u-v)\s(u+v)\s(2u+w)\s(\l_{1k}w-w)\s(\l_{1k}w+w)},\label{B-6}
\eea to derive the last term on the right side of the above
equation. The equation (\ref{B-6}) is a consequence of the
identity (\ref{identity}). Again using the identity
(\ref{identity}), after some long tedious calculation, we finally
obtain the commutation relation (\ref{Rel-2}) from the above
exchange relation by noting the definitions (\ref{Def-AB}) and
(\ref{Def-D}).

For $i_0=j_0=1$, $i_3=i\neq 1$ and $j_3=j\neq 1$, we obtain \bea
&&B_i(\l+\hat{i}+\hat{1}|u) B_j(\l+\hat{i}+\hat{j}|v)=
\sum_{\b,\a=2}^n\frac{R(u-v,\l)^{\b\,\a}_{j\,i}
R(u+v,\l)^{1\,\b}_{1\,\b}} {R(u-v,\l)^{1\,1}_{1\,1}
R(u+v,\l)^{1\,i}_{1\,i}} \no\\
&&~~~~~~~~~~~~~~~~~~~~~~~~~~~~~~~~~~~~~~~~~~~~~~\times
B_{\b}(\l+\hat{1}+\hat{\b}|v)B_{\a}(\l+\hat{i}+\hat{j}|u),\eea
which leads to the commutation relation (\ref{Rel-3}).

\section*{Appendix C: The action of $\T^i_j$ on the pseudo-vacuum state}
\setcounter{equation}{0}
\renewcommand{\theequation}{C.\arabic{equation}}

Using the same method  as applied for the calculation of the
actions of $\T^1_1$ (\ref{TF-V1}) and $\T^i_1$ (\ref{TF-V2}) on
the pseudo-vacuum state (\ref{Vac}) in subsection 4.2, we have
\bea
&&\T(\l_0-N\hat{1},\l_0|u;\xi)^i_j|vac\rangle^{\l_0-N\hat{1}}_{\l_0}=
k(\l_0|u;\xi)_1T(\l_0-N\hat{1}-\hat{j},\l_0-\hat{1}|u)^i_1\no\\
&&~~~~~~~~~~\times S(\l_0-N\hat{1},\l_0|u)^1_j
|vac\rangle^{\l_0-N\hat{1}}_{\l_0}\no\\
&&~~~~~~+\d^i_jk(\l_0|u;\xi)_j\prod_{k=1}^N
R(u-z_k,\l_0-(N-k)\hat{1})^{1j}_{1j}
R(u+z_k,\l_0-(N-k)\hat{1})^{j1}_{j1}\no\\
&&~~~~~~~~~~\times
|vac\rangle^{\l_0-N\hat{1}}_{\l_0},~~i,j=2,\cdots,n.\label{C-1}\eea
The first term on the right hand of the above equation cannot be
calculated directly by the same method. However, we can calculate
the first term by the following way.

We can derive the following exchange relations from ``RLL"
relation (\ref{Relation1}) \bea
T_1(u)R^B_{12}(2u)T^{-1}_2(-u)=T^{-1}_2(-u)R_{12}^B(2u)T_1(u).
\eea Multiplying both sides of the  above equation from the left
by
$\tilde{\phi}_{\l_0-N\hat{1}-\hat{i}+\hat{j},\l_0-N\hat{1}-\hat{i}}(u)
\otimes\bar{\phi}_{\l_0+\hat{1},\l_0}(-u)$ and from the right by
$\phi_{\l_0+\hat{1},\l_0}(u)\otimes
\phi_{\l_0-N\hat{1},\l_0-N\hat{1}-\hat{i}}(-u)$, we obtain  the
following exchange relation from the face-vertex correspondence
relations (\ref{Face-vertex}) and (\ref{Face-vertex1})-
(\ref{Face-vertex4}) \bea
&&\sum_{\a=1}^nR(2u,\l_0+\hat{1})^{\a\,1}_{1\,\a}T(\l_0-N\hat{1}-\hat{i},
\l_0-\hat{\a}|u)^j_{\a}S(\l_0-N\hat{1}, \l_0|u)^{\a}_i\no\\
&&~~~~~~=\sum_{\a,\b=1}^n
R(2u,\l_0-N\hat{1}+\hat{\a})^{j\,\b}_{\a\,i}S(\l_0-N\hat{1}+\hat{\a},
\l_0+\hat{1}|u)^1_{\b}T(\l_0-N\hat{1}, \l_0|u)^{\a}_1.\no\\
\eea Acting both sides on the pseudo-vacuum state
$|vac\rangle^{\l_0-N\hat{1}}_{\l_0}$, and using the equations
(\ref{Action-1})-(\ref{Action-2}), we obtain \bea
&&T(\l_0-N\hat{1}-\hat{i},\l_0-\hat{1}|u)^j_1S(\l_0-N\hat{1},\l_0|u)^1_i
|vac\rangle^{\l_0-N\hat{1}}_{\l_0}=\d^j_i\lt\{
R(2u,\l_0-(N-1)\hat{1})^{j1}_{1j}\rt.\no\\
&&~~~~~~\lt.-R(2u,\l_0+\hat{1})^{j1}_{1j}
\prod_{k=1}^NR(u-z_k,\l_0-(N-k)\hat{1})^{1j}_{1j}
R(u+z_k,\l_0-(N-k)\hat{1})^{j1}_{j1}\rt\}\no\\
&&~~~~~~~~\times|vac\rangle^{\l_0-N\hat{1}}_{\l_0}.\label{C-2}\eea
The equation (\ref{TF-V3}) is a simple consequence of the
equations (\ref{C-1}) and (\ref{C-2}).


\end{document}